\documentclass{article}

\usepackage[square,numbers]{natbib}

\usepackage{hyperref}
\usepackage{a4wide}
\usepackage{amsmath}
\usepackage{amssymb}
\usepackage{amsthm}
\usepackage{dsfont}
\usepackage{graphicx}
\usepackage{comment}
\usepackage[utf8]{inputenc}
\usepackage{tikz}
\usetikzlibrary{calc, positioning, fit, shapes.misc}

\newtheorem{theorem}{Theorem}
\newtheorem{lemma}[theorem]{Lemma}
\newtheorem{corollary}[theorem]{Corollary}
\newtheorem{claim}[theorem]{Claim}
\newtheorem{proposition}[theorem]{Proposition}
\newtheorem{experiment}[theorem]{Random Experiment}
\newtheorem{remark}[theorem]{Remark}

\newtheorem{innercustomthm}{Lemma}
\newenvironment{customthm}[1]
  {\renewcommand\theinnercustomthm{#1}\innercustomthm}
  {\endinnercustomthm}

\newtheorem{innercustomcla}{Claim}
\newenvironment{customcla}[1]
  {\renewcommand\theinnercustomcla{#1}\innercustomcla}
  {\endinnercustomcla}

\newcommand*{\email}[1]{%
    \href{mailto:#1}{#1}\par
    }

\newcommand{\OPT}{\mathrm{OPT}}
\newcommand{\E}{\mathbb{E}}
\renewcommand{\P}{\mathbb{P}}
\newcommand{\C}{\mathcal{C}}
\newcommand{\K}{\mathcal{K}}
\newcommand{\I}{R}
\newcommand{\F}{\mathcal{F}}
\newcommand{\N}{\mathcal{N}}
\newcommand{\s}{\mathcal{S}}

\usepackage[colorinlistoftodos]{todonotes}

\title{The Submodular Santa Claus Problem \\ in the Restricted Assignment Case\footnote{This research was supported by the Swiss National Science Foundation project
200021-184656 “Randomness in Problem Instances and Randomized Algorithms.”}}

\author{Etienne Bamas\footnote{EPFL, Switzerland, \email{etienne.bamas@epfl.ch}} \and Paritosh Garg\footnote{EPFL, Switzerland, \email{paritosh.garg@epfl.ch}} \and Lars Rohwedder\footnote{EPFL, Switzerland, \email{lars.rohwedder@epfl.ch}}}

\date{}

\begin{document}

\maketitle

\begin{abstract}
The submodular Santa Claus problem
was introduced in
a seminal work by Goemans, Harvey, Iwata, and Mirrokni (SODA'09) as
an application of their structural result.
In the mentioned problem $n$ unsplittable resources have to
be assigned to $m$ players,
each with a monotone submodular utility
function $f_i$.
The goal is to maximize
$\min_i f_i(S_i)$ where
$S_1,\dotsc,S_m$ is a partition of
the resources.
The result by Goemans et al. implies
a polynomial time $O(n^{1/2 +\varepsilon})$-approximation algorithm.

Since then progress on this problem was
limited to the linear case, that is,
all $f_i$ are linear functions.
In particular, a line of research has
shown that there is a polynomial time
constant approximation algorithm
for linear valuation functions in the
restricted assignment case.
This is the special case
where each player is given a set of
desired resources $\Gamma_i$ and
the individual valuation functions
are defined as $f_i(S) = f(S \cap \Gamma_i)$
for a global linear function $f$.
This can also be interpreted as maximizing
$\min_i f(S_i)$ with additional assignment
restrictions, i.e., resources can only
be assigned to certain players.

In this paper we make comparable
progress for the submodular variant.
Namely, if $f$ is a monotone submodular
function, we can in polynomial time
compute an $O(\log\log(n))$-approximate
solution.

    
\end{abstract}
\pagebreak

\section{Introduction}
In the Santa Claus problem (sometimes referred to as Max-Min Fair Allocation) we are given a set of $n$ players $P$ and a set of $m$ indivisible resources $R$.
In its full generality, each player $i\in P$ has a utility function $f_i:2^R\mapsto \mathbb{R}_{\ge 0}$, where $f_i(S)$ measures the
happiness of player $i$ if he is assigned the resource set $S$.
The goal is to find a partition of the resources that maximizes the happiness of the least happy player. Formally, we want to find a partition  $\{S_i\}_{i\in P}$ of the resources that maximizes 
\begin{equation*}
    \min_{i\in P} f_i(S_i) .
\end{equation*}
Most of the recent literature on this problem focuses on 
cases where $f_i$ is a linear function for all players $i$.
If we assume all valuation functions are linear, 
the best approximation algorithm known for this problem, 
designed by Chakrabarty, Chuzhoy, and Khanna~\cite{DBLP:conf/focs/ChakrabartyCK09}, has an
approximation rate of
$n^{\epsilon}$ and runs in time $n^{O(1/\epsilon)}$ for $\epsilon\in\Omega(\log\log(n)/\log(n))$.
On the negative side,
it is only known that computing a $(2 - \delta)$-approximation is NP-hard~\cite{LenstraSchmoysTardos}.
Apart from this there has been significant attention on the so-called 
\emph{restricted assignment case}.
Here the utility functions are defined by one linear function $f$ and a set of resources $\Gamma_i$ for each player $i$.
Intuitively, player $i$ is interested in the resources $\Gamma_i$,
whereas the other resources are worthless for him.
The individual utility functions are then implicitly defined by
$f_i(S)=f(S\cap \Gamma_i)$. In a seminal work Bansal and Srividenko~\cite{BansalSrividenko} provide a $O(\log \log (m)/\log \log \log (m))$-approximation algorithm for this case.
This was improved by Feige~\cite{Feige} to an $O(1)$-approximation.
Further progress on the constant or the running time was made since then,
see e.g.~\cite{DBLP:journals/talg/AnnamalaiKS17, DBLP:conf/soda/DaviesRZ20, DBLP:conf/icalp/ChengM19, DBLP:conf/icalp/ChengM18, JANSEN2020106025, Asadpour_local_search, Polacek}.

Let us now move to the non-linear case.
Indeed, the problem becomes hopelessly difficult without any restrictions
on the utility functions.
Consider the following reduction from set packing. There are sets of resources $\{S_1,\dotsc,S_k\}$ and all utility functions are equal and
defined by $f_i(S) = 1$ if $S_j \subseteq S$ for some $j$
and $f_i(S) = 0$ otherwise. Deciding whether there are $m$ disjoint sets in $S_1,\dotsc,S_k$ (a classical NP-hard problem) 
is equivalent to deciding whether the optimum of the Santa Claus
problem is non-zero. In particular, obtaining any bounded approximation ratio for Santa Claus in this case is NP-hard.


Two naturally arising properties of utility functions
are monotonicity and submodularity,
see for example the related submodular welfare problem~\cite{DBLP:journals/geb/LehmannLN06,DBLP:conf/stoc/Vondrak08}
where the goal is to maximize
$\sum_i f_i(S_i)$.
A function $f$ is monotone, if $f(S) \le f(T)$ for all $S\subseteq T$.
It is submodular, if $f(S\cup \{a\}) - f(S) \ge f(T\cup\{a\}) - f(T)$ for all $S\subseteq T$ and $a\notin T$.
The latter is also known as the \emph{diminishing returns} property in economics.
A standard assumption on monotone submodular functions (used throughout this work) is that the value on the empty set is zero, i.e., $f(\emptyset) = 0$. 
Goemans, Harvey, Iwata, and Mirrokni~\cite{goemans2009approximating} first considered the Santa Claus
problem with monotone submodular utility functions as an application
of their fundamental result on submodular functions.
Together with the algorithm of~\cite{DBLP:conf/focs/ChakrabartyCK09}
it implies an $O(n^{1/2+\epsilon})$-approximation in time $O(n^{1/\epsilon})$.

In this paper we investigate the restricted assignment case with
a monotone submodular utility function. That is,
all utility functions are defined by $f_i(S)=f(S\cap \Gamma_i)$,
where $f$ is a monotone submodular function and $\Gamma_i$ is 
a subset of resources for each players $i$. Before our work, the state-of-the-art for this problem was the $O(n^{1/2+\epsilon})$-approximation algorithm mentioned above, since none of the previous results for the restricted assignment case with a linear utility function apply when the utility function becomes monotone submodular.

\subsection{Overview of results and techniques}

Our main result is an approximation algorithm for
the submodular Santa Claus problem in the restricted assignment case.

\begin{theorem}
\label{thm:main}
There is a randomized polynomial time $O(\log \log (n))$-approximation algorithm for the restricted assignment case with a monotone submodular utility function.
\end{theorem}

Our way to this result is organised as follows. In Section~\ref{sec:reduction to hypergraph}, we first reduce our problem to a hypergraph matching problem (see
next paragraph for a formal definition).
We then solve this problem using Lovasz Local Lemma (LLL) in Section~\ref{sec:hypergraph problem}.
In~\cite{BansalSrividenko} the authors also reduce to a hypergraph matching problem which they then solve using LLL,
although both parts are substantially simpler.
The higher generality of our utility functions is
reflected in the more general hypergraph matching problem.
Namely, our problem is precisely the weighted variant
of the (unweighted) problem in~\cite{BansalSrividenko}.
We will elaborate later in this section why the previous
techniques do not easily extend to the weighted variant.


\paragraph{The hypergraph matching problem.} After the reduction in Section \ref{sec:reduction to hypergraph} we arrive at the following problem. There is a hypergraph
$\mathcal H = (P\cup R, \mathcal C)$ with
hyperedges $\mathcal C$ over the vertices $P$ and $R$.
We write $m = |P|$ and $n = |R|$.
We will refer to hyperedges as configurations, the
vertices in $P$ as players and $R$ as resources%
\footnote{We note that these do not have to be the same players and resources as in the Santa Claus problem we reduced from, but $n$ and $m$ do not increase.}.
Moreover, a hypergraph is said to be regular if all vertices in $P$ and $R$ have the same degree, that is,
they are contained in the same number of configurations.

The hypergraph may contain multiple copies of the same configuration.
Each configuration $C\in\mathcal C$ contains exactly one vertex in $P$, that is, $|C\cap P| = 1$.
Additionally, for each configuration $C\in \C$ the resources $j\in C$ have weights $w_{j,C} \ge 0$.
We emphasize
that the same resource $j$ can be given different weights in two different configurations, that is,
we may have $w_{j,C}\neq w_{j,C'}$ for two different configurations $C,C'$.

We require to select for each player $i\in P$ one configuration $C$ that contains $i$. For each configuration $C$ that was selected we require to assign a subset of the resources in $C$
which has a total weight of at least $(1/\alpha) \cdot \sum_{j\in C} w_{j,C}$ to the player in $C$.
A resource can only be assigned to one player. We call such a solution an $\alpha$-relaxed perfect matching. One seeks to minimize $\alpha$.

We show that every regular hypergraph has an $\alpha$-relaxed perfect matching for some $\alpha=O(\log \log (n))$ assuming that $w_{j,C}\leq (1/\alpha) \cdot \sum_{j'\in C} w_{j',C}$ for all $j,C$, that is, all weights are small compared to
the total weight of the configuration.
Moreover, we can find such a matching in randomized polynomial time. In the reduction we use this result
to round a certain LP relaxation and $\alpha$ essentially
translates to the approximation rate.
This result generalizes that of Bansal and Srividenko on hypergraph matching in the following way. They proved the same result for unit weights and uniform hyperedges, that is,
$w_{j,C}=1$ for all $j,C$ and all hyperedges have the same number of resources\footnote{In fact they get a slightly better ratio
of $\alpha = O(\log\log(m) / \log\log\log(m))$.}.
In the next paragraph we briefly go over the techniques
to prove our result for the hypergraph matching problem.

\paragraph{Our techniques.}
Already the extension from uniform to non-uniform
hypergraphs (assuming unit weights)
is highly non-trivial and captures the
core difficulty of our result.
Indeed, we show with a (perhaps surprising) reduction,
that we can reduce our weighted hypergraph matching problem to the unweighted (but non-uniform) version by introducing some bounded dependencies between the choices of the different players.
For sake of brevity we therefore focus in this section on
the unweighted non-uniform variant, that is,
we need to assign to each player a configuration $C$ and
at least $|C| / \alpha$ resources in $C$.
We show that for any regular hypergraph there exists such a matching for $\alpha = O(\log \log (n))$ assuming that all configurations contain at least $\alpha$ resources and we can find it in randomized polynomial time.
Without the assumption of uniformity the problem becomes significantly more challenging.
To see this, we lay out the techniques of Bansal and Srividenko that allowed them to solve the problem in the uniform case. We note that for $\alpha = O(\log(n))$ the statement is
easy to prove: We select for each player $i$
one of the configurations containing $i$ uniformly
at random. Then by standard concentration bounds each
resource is contained in at most $O(\log(n))$
of the selected configurations with high probability.
This implies that there is a fractional assignment of resources to configurations such that each of the
selected configurations $C$ receives
$\lfloor |C| / O(\log(n)) \rfloor$ of the resources
in $C$. By integrality of the bipartite matching polytope, there
is also an integral assignment with this property.

To improve to $\alpha= O(\log \log (n))$ in the uniform case, Bansal and Srividenko proceed as follows. Let $k$ be the size of each configuration. First they reduce the degree
of each player and resource to $O(\log(n))$ using
the argument above, but taking $O(\log(n))$ configurations for each player.
Then they sample
uniformly at random $O(n \log(n) / k)$ resources
and drop all others. This is sensible, because
they manage to prove the (perhaps surprising) fact that an
$\alpha$-relaxed perfect matching with respect to
the smaller set of resources is still an
$O(\alpha)$-relaxed perfect matching with respect
to all resources with high probability (when assigning the dropped resources
to the selected configurations appropriately).
Indeed, the smaller instance is easier to solve:
With high probability all configurations have size $O(\log(n))$ and this greatly reduces the dependencies
between the bad events of the random experiment above
(the event that a resource is contained
in too many selected configurations).
This allows them to apply Lov\'asz Local Lemma (LLL)
in order to show that with positive probability the
experiment succeeds for $\alpha = O(\log\log(n))$.

It is not obvious how to extend this approach
to non-uniform hypergraphs:
Sampling a fixed fraction of the
resources will either make the small configurations
empty---which makes it impossible to retain guarantees for the original instance---or it leaves the big configurations big%
---which fails to reduce the dependencies enough to
apply LLL. Hence it requires new sophisticated ideas
for non-uniform hypergraphs, which we describe next.

Suppose we are able to find a set $\mathcal K\subseteq \mathcal C$ of configurations (one for each player)
such that for each $K\in\mathcal K$ the sum of intersections $|K\cap K'|$ with smaller configurations $K'\in \mathcal K$ is
very small, say at most $|K| / 2$.
Then it is easy to derive a $2$-relaxed perfect matching: We iterate over all $K\in\mathcal K$ from large to small and reassign all resources to $K$
(possibly stealing them from the configuration that previously had them).
In this process every configuration gets stolen at most $|K| / 2$ of its resources,
in particular, it keeps the other half.
However, it is non-trivial to obtain a property like the one mentioned above. If we take a random configuration for each player, the
dependencies of the intersections are too complex.
To avoid this we invoke an advanced variant of the sampling approach where we construct not only one
set of resources, but a hierarchy of
resource sets $\I_0\supseteq \cdots \supseteq \I_d$ by repeatedly dropping a fraction of resources from the previous set.
We then formulate bad events based on the intersections
of a configuration $C$ with smaller configurations $C'$,
but we write it only considering a resource set $\I_k$ of convenient granularity
(chosen based on the size of $C'$).
In this way we formulate a number of bad events using
various sets $\I_k$. This succeeds in reducing
the dependencies enough to apply LLL.
Unfortunately, even with this new way of defining bad events, the guarantee that for each $K\in\mathcal K$ the sum of intersections $|K\cap K'|$ with smaller configurations $K'\in \mathcal K$ is at most $|K| / 2$ is still too much to ask. We can only prove some weaker property which makes it more difficult to reconstruct a good solution from it. The reconstruction still starts from the biggest configurations and iterates to finish by including the smallest configurations but it requires a delicate induction where at each step, both the resource set expands and some new small configurations that were not considered before come into play.

\paragraph{Additional implications of non-uniform hypergraph matchings to the Santa Claus problem.} We believe this hypergraph matching problem is interesting in its own right. Our last contribution is to show that finding good matchings in unweighted hypergraphs with fewer assumptions than ours would have important applications for the Santa Claus problem with linear utility functions. We recall that here, each player $i$ has its own utility function $f_i$ that can be any linear function. In this case, the best approximation algorithm is due to Chakrabarty, Chuzhoy, and Khanna~\cite{DBLP:conf/focs/ChakrabartyCK09} who gave a $O(n^{\epsilon})$-approximation running in time $O(n^{1/\epsilon})$. In particular, no sub-polynomial approximation running in polynomial time is known. Consider as before $\mathcal H = (P\cup R, \mathcal C)$ a non-uniform hypergraph with unit weights ($w_{j,C}=1$ for all $j,C$ such that $j\in C$). Finding the smallest $\alpha$ (or an approximation of it) such that there exists an $\alpha$-relaxed perfect matching in $\mathcal H$ is already a very non-trivial question to solve in polynomial time.

We show, via a reduction, that a $c$-approximation for this problem would yield a $O((c\log^* (n))^2)$-approximation for the Santa Claus problem with arbitrary linear utility functions. In particular, any sub-polynomial approximation for this problem would significantly improve the state-of-the-art\footnote{We mention that our result on relaxed matchings in Section \ref{sec:hypergraph problem} does not imply an $O(\log \log (n))$-approximation for this problem since we make additional assumptions on the regularity of the hypergraph or the size of hyperedges.}. All the details of this last result can be found in Section \ref{sec:reduction santa claus}.

\paragraph{A remark on local search techniques.} We focus here on an extension of the LLL technique of Bansal and Srividenko. However, another technique proved itself very successful for the Santa Claus problem in the restricted assignment case with a linear utility function. 
This is a local search technique discovered by Asadpour, Feige, and Saberi~\cite{Asadpour_local_search} who used it to give a non-constructive proof that the integrality gap of the configuration LP of Bansal and Srividenko is at most $4$. One can wonder if this technique could also be extended to the submodular case as we did with LLL. Unfortunately, this seems problematic as the local search arguments heavily rely on amortizing different volumes of configurations (i.e., the sum of
their resources' weights or the number of resources 
in the unweighted case).
Amortizing the volumes of configurations works well, if each configuration
has the same volume, which is the case for the problem
derived from linear valuation functions, but not the one derived
from submodular functions.
If the volumes differ then these amortization arguments break and
the authors of this paper believe this is a fundamental problem for
generalizing those arguments.


\section{Reduction to hypergraph matching problem}
\label{sec:reduction to hypergraph}
In this section we give a reduction of the restricted submodular
Santa Claus problem to the hypergraph matching problem.
As a starting point we solve the configuration LP,
a linear programming relaxation of our problem.
The LP is constructed using a parameter $T$ which
denotes the value of its solution.
The goal is to find the maximal $T$ such
that the LP is feasible.
In the LP we have a variable $x_{i,C}$ for every player $i\in P$ and
every configuration $C\in \C(i, T)$.
The configurations $\C(i, T)$ are defined as the sets of resources $C \subseteq \Gamma_i$ such that $f(C) \ge T$.
We require every player $i\in P$ to have at least one configuration
and every resource $j \in \I$ to be contained in at most one configuration.
\begin{align*}
     \sum_{C\in \C(i, T)} & x_{i,C} \ge 1 \quad \text{ for all } i \in P \\
     \sum_{i\in P}\sum_{C\in \C(i,T) : j \in C} & x_{i,C} \leq 1 \quad \text{ for all } j \in \I \\
    &  x_{i,C} \geq 0 \quad \text{ for all } i\in P, C \in \C(i, T)
\end{align*}
Since this linear program has exponentially many variables,
we cannot directly solve it in polynomial time.
We will give a polynomial time constant approximation for it
via its dual. This is similar to the linear variant
in~\cite{BansalSrividenko}, but
requires some more work. In their case they can reduce the
problem to one where the separation problem of the dual can
be solved in polynomial time. In our case even the separation
problem can only be approximated. Nevertheless, this is sufficient
to approximate the linear program in polynomial time.

\begin{theorem}\label{thm:config-LP}
  The configuration LP of the restricted submodular Santa Claus
  problem can be approximated within a factor of $(1 - 1/e)/2$
  in polynomial time.
\end{theorem}
We defer the proof of this theorem to Appendix~\ref{appendix_lp}.
Given a solution $x^*$ of the configuration LP we want
to arrive at the
hypergraph matching problem from the introduction such
that an $\alpha$-relaxed perfect matching of that
problem corresponds to an $O(\alpha)$-approximate
solution of the restricted submodular Santa Claus problem.
Let $T^*$ denote the value of the solution $x^*$.
We will define a resource $j\in R$ as \emph{fat} if 
\begin{equation*}
    f(\{j\}) \geq \frac{T^*}{100 \alpha} .
\end{equation*}
Resources that are not fat are called \emph{thin}.
We call a configuration $C\in\C(i, T)$ thin,
if it contains only thin resources and
denote by $\C_t(i,T) \subseteq \C(i, T)$ the
set of thin configurations.
Intuitively in order to obtain an
$O(\alpha)$-approximate solution, it suffices
to give each player $i$ either one fat resource
$j\in \Gamma_i$ or
a thin configuration $C\in\C_t(i,T^*/O(\alpha))$. For our
next step towards the hypergraph problem we use
a technique borrowed from Bansal and Srividenko~\cite{BansalSrividenko}.
This technique allows us to simplify the structure of the problem significantly
using the solution of the configuration LP.
Namely, one can find a partition of the players into clusters such that
we only need to cover one player from each cluster with thin resources.
All other players can then be covered by fat resources.
Informally speaking, the following lemma is proved by sampling configurations
randomly according to a distribution derived in a non-trivial way
from the configuration LP.
\begin{lemma}\label{lem:config-sample}
  Let $\ell \ge 12\log (n)$.
  Given a solution of value $T^*$ for the configuration LP
  in randomized polynomial time we can find a partition of the players into clusters $K_1\cup\cdots \cup K_k\cup Q = P$ and multisets of configurations
  $\C_h \subseteq \bigcup_{i\in K_h} \C_T(i, T^*/5)$, $h=1,\dotsc,k$, such that
  \begin{enumerate}
      \item $|\C_h| = \ell$ for all $h=1,\dotsc,k$ and
      \item Each small resource appears in at most $\ell$ configurations of $\bigcup_h \C_h$.
    \item given any $i_1\in K_1, i_2\in K_2,\dotsc,i_k\in K_k$
    there is a matching of fat resources to players
    $P\setminus\{i_1,\dotsc,i_k\}$ such that each of these players $i$ gets a unique fat resource $j\in\Gamma_i$.
  \end{enumerate}
\end{lemma}
The role of the players $Q$ in the lemma above is that each one of them gets
a fat resource for certain.
The proof follows closely that in~\cite{BansalSrividenko}. For completeness we
include it in Appendix~\ref{appendix_lp}.
We are now ready to define
the hypergraph matching instance.
The vertices of our hypergraph are the clusters
$K_1,\dotsc,K_k$ and the thin resources.
Let $\C_1,\dotsc,\C_k$ be the multisets of configurations
as in Lemma~\ref{lem:config-sample}.
For each $K_h$ and $C\in\C_h$ there is a
hyperedge containing $K_h$ and all resources in $C$.
Let $\{j_1,\dotsc,j_\ell\} = C$ ordered arbitrarily,
but consistently. Then we define the weights
as normalized marginal gains of resources if they are
taken in this order, that is,
\begin{equation*}
    w_{j_i, C} = \frac{5}{T^*} f(\{j_i\} \mid \{j_1,\dotsc,j_{i-1}\}) = \frac{5}{T^*}  (f(\{j_1,\dotsc,j_{i-1}, j_i\})-f(\{j_1,\dotsc,j_{i-1}\})).
\end{equation*}
This implies that $\sum_{j\in C} w_{j, C} \ge 5 f(C) / T^* \ge 1$ for each $C\in\C_h$, $h=1,\dotsc,k$.
\begin{lemma}
  Given an $\alpha$-relaxed perfect matching to the
  instance as described by the reduction, one can
  find in polynomial time an $O(\alpha)$-approximation
  to the instance of restricted submodular Santa Claus.
\end{lemma}
\begin{proof}
  The $\alpha$-relaxed perfect matching implies that
  cluster $K_h$ gets some small resources $C'$
  where $C'\subseteq C$ for some $C\in\C_h$
  and $\sum_{j\in C'} w_{j, C} \ge 1/\alpha$.
  By submodularity we have that
  $f(C') \ge T^* / (5 \alpha)$.
  Therefore we can satisfy one player in each
  cluster using thin resources and by
  Lemma~\ref{lem:clusters} all others using
  fat resources.
\end{proof}
The proof above is the most critical place in the paper where we make use of the
submodularity of the valuation function $f$.
We note that since all resources considered are thin resources we have, by submodularity of $f$, the assumption that 
\begin{equation*}
    w_{j,C} \leq \frac{5}{T^*}f(\{j\}) \leq  \frac{5}{T^*}\frac{T^*}{100\alpha} \leq \frac{5}{100\alpha} \sum_{j\in C} w_{j,C}
\end{equation*} for all $j,C$ such that $j\in C$. This means that the weights are all small enough, as promised in introduction. From now on, we will assume that $\sum_{j\in C} w_{j,C}=1$ for all configurations $C$. This is w.l.o.g. since we can just rescale the weights inside each configuration. This does not hurt the property that all weights are small enough.

\subsection{Reduction to unweighted hypergraph matching}
Before proceeding to the solution of this hypergraph matching problem, we first give a reduction to an unweighted variant of the problem. We will then solve this unweighted variant in the next section. First, we note that we can assume that all the weights $w_{j,C}$ are powers of $2$ by standard rounding arguments. This only loses a constant factor in the approximation rate. Second, we can assume that inside each configuration $C$, each resource has a weight that is at least a $1/(2n)$. Formally, we can assume that 
\begin{equation*}
    \min_{j\in C}w_{j,C}\geq 1/(2n)
\end{equation*}
for all $C\in \C$.
If this is not the case for some $C\in \C$, simply delete from $C$ all the resources that have a weight less than $1/(2n)$. By doing this, the total weight of $C$ is only decreased by a factor $1/2$ since it looses in total at most a weight of
\begin{equation*}
    n\cdot \frac{1}{2n} = \frac{1}{2}.
\end{equation*} (Recall that we rescaled the weights so that $\sum_{j\in C} w_{j,C}=1$).

Hence after these two operations, an $\alpha$-relaxed perfect matching in the new hypergraph is still an $O(\alpha)$-relaxed perfect matching in the original hypergraph. From there we reduce to an unweighted variant of the matching problem. Note that each configuration contains resources of at most $\log (n)$ different possible weights (powers of $2$ from $1 /(2n)$ to $1 / \alpha$). We create the following new unweighted hypergraph $\mathcal H'=(P'\cup R,\mathcal C')$. The resource set $R$ remains unchanged. For each player $i\in P$, we create $\log (n)$ players, which
later correspond each to a distinct weight. We will say that the players obtained from duplicating the original player form a \textit{group}. For every configuration $C$ containing player $i$ in the hypergraph $\mathcal H$, we add a set $\s_C=\{C_1,\ldots ,C_s, \ldots ,C_{\log(n)}\}$ of configurations in $\mathcal H'$. $C_s$ contains player $i_s$ and all resources that are given a weight $2^{-(s+1)}$ in $C$. In this new hypergraph, the resources are not weighted. Note that if the hypergraph $\mathcal H$ is regular then $\mathcal H'$ is regular as well.

Additionally, for a group of player and a set of $\log(n)$ configurations (one for each player in the group), we say that this set of configurations is \textit{consistent} if all the configurations selected are obtained from the same configuration in the original hypergraph $\mathcal H$ (i.e. the selected configurations all belong to $\s_C$ for some $C$ in $\mathcal H$).

Formally, we focus of the following problem. Given the regular hypergraph $\mathcal H'$, we want to select, for each group of $\log (n)$ players, a consistent set of configurations $C_1,\ldots, C_s, \ldots ,C_{\log(n)}$ and assign to each player $i_s$ a subset of the resources in the corresponding configuration $C_s$ so that $i_s$ is assigned at least $\left\lfloor |C_s|/\alpha\right\rfloor$ resources. No resource can be assigned to more than one player. We refer to this assignment as a consistent $\alpha$-relaxed perfect matching. Note that in the case where $|C_s|$ is small (e.g. of constant size) we are not required to assign any resource to player $i_s$.

\begin{lemma}
A consistent $\alpha$-relaxed matching in $\mathcal H'$ induces a $O(\alpha)$-relaxed matching in $\mathcal H$.
\end{lemma}
\begin{proof}
Let us consider a group of $\log (n)$ players $i_1,\ldots , i_s, \ldots ,i_{\log (n)}$ in $\mathcal H'$ corresponding to a player $i$ in $\mathcal H$. These players are assigned a consistent set of configurations $C_1,\ldots , C_s, \ldots, C_{\log (n)}$ that correspond to a partition of a configuration in $\mathcal H$. Moreover, each player $i_s$ is assigned $\left\lfloor |C_s|/\alpha\right\rfloor$ resources from $C_s$. We have two cases. If $|C_s|\geq \alpha$ then we have that $i_s$ is assigned at least
\begin{equation*}
    \left\lfloor |C_s|/\alpha\right\rfloor\geq |C_s|/(2\alpha)
\end{equation*} resources from $C_s$.
On the other hand, if $\left\lfloor |C_s|/\alpha\right\rfloor=0$ 
then the player $i_s$ might not be assigned anything. However, we claim that that the configurations $C_s$ of cardinality less than $\alpha$ can represent at most a $1/5$ fraction of the total weight of the configuration $C$ in the original weighted hypergraph. To see this note that the total weight they represent is upper bounded by
\begin{equation*}
    \alpha \left(\sum_{k=\log(100\alpha/5)}^{\infty} \frac{1}{2^k}\right) = \alpha\left(\frac{5}{100\alpha}\sum_{k=0}^{\infty} \frac{1}{2^k}\right) \leq \frac{10}{100} =  \frac{1}{10}\sum_{j\in C}w_{j,C}.
\end{equation*}

Hence, the consistent $\alpha$-relaxed matching in $\mathcal H'$ induces in a straightforward way a matching in $\mathcal H$ where every player gets at least a fraction $1/(2\alpha) \cdot (1-1/10) \geq 1/(3\alpha)$ of the total weight of the appropriate configuration. This means that the consistent $\alpha$-relaxed perfect matching in $\mathcal H'$ is indeed a $(3\alpha)$-relaxed perfect matching in $\mathcal H$.
\end{proof}

\section{Matchings in regular hypergraphs}
\label{sec:hypergraph problem}

In this section we solve the hypergraph matching problem we arrived to in the previous section. For convenience, we give a self contained definition of the problem before formulating and proving our result. 

\paragraph{Input:} We are given $\mathcal H = (P\cup R, \mathcal C)$ a hypergraph with
hyperedges $\mathcal C$ over the vertices $P$ (players) and $R$ (resources) with
$m = |P|$ and $n = |R|$. As in previous sections, we will refer to hyperedges as configurations. 
Each configuration $C\in\mathcal C$ contains exactly one vertex in $P$,
that is, $|C\cap P| = 1$. The set of players is partitioned into groups of size at most $\log (n)$, we will use $A$ to denote a group. These groups are disjoint and contain all players. Finally there exists an integer $\ell$ such that for each group $A$ there are $\ell$ consistent sets of configurations. A consistent set of configurations for a group $A$ is a set of $|A|$ configurations such that all players in the group appear in exactly one of these configurations. We will denote by $\s_A$ such a set and for a player $i\in A$, we will denote by $\s_A^{(i)}$ the unique configuration in $\s_A$ containing $i$. Finally, no resource appears in more than $\ell$ configurations.
We say that the hypergraph is regular (although some resources may appear in less than
$\ell$ configurations).

\paragraph{Output:} We wish to select a matching that covers all players in $P$. More precisely, for each group $A$ we want to select a consistent set of configurations (denoted by $\{\s_A^{(i)}\}_{i\in A}$). Then for each player $i\in A$, we wish to assign a subset of the resources in $\s_A^{(i)}$ to the player $i$ such that:
\begin{enumerate}
    \item No resource is assigned to more than one player in total.
    \item For any group $A$ and any player $i\in A$, player $i$ is assigned at least 
    \begin{equation*}
        \left\lfloor \frac{\s_A^{(i)}}{\alpha}\right\rfloor
    \end{equation*} resources from $\s_A^{(i)}$.
\end{enumerate}
We call this a consistent $\alpha$-relaxed perfect matching. Our goal in this section will be to prove the following theorem.

\begin{theorem}\label{thm:unweighted_hypergraph}
Let $\mathcal H=(P\cup R, \mathcal C)$ be a regular (non-uniform) hypergraph where the set of players is partitioned into groups of size at most $\log (n)$. Then we can, in randomized polynomial time, compute a consistent $\alpha$-relaxed perfect matching for $\alpha=O(\log \log (n))$.
\end{theorem}

We note that Theorem \ref{thm:unweighted_hypergraph} together with the reduction from the previous section will prove our main result (Theorem \ref{thm:main}) stated in introduction.

\subsection{Overview and notations}

To prove Theorem~\ref{thm:unweighted_hypergraph}, we introduce the following notations. 
Let $\ell \in \mathbb N$ be the regularity parameter as described in the problem input (i.e. each group has $\ell$ consistent sets and each resource appears in no more than $\ell$ configurations). As we proved in Lemma \ref{lem:config-sample} we can assume with standard sampling arguments that $\ell =300.000\log^{3}(n)$ at a constant loss. If this is not the case because we might want to solve the hypergraph matching problem by itself (i.e. not obtained by the reduction in Section \ref{sec:reduction to hypergraph}), the proof of Lemma \ref{lem:config-sample} can be repeated in a very similar way here.


For a configuration $C$, its size will be defined as $|C\cap R|$ (i.e. its cardinality over the resource set). For each player $i$, we denote by $\C_i$ the set of configurations that contain $i$. We now group the configurations in $\C_i$ by size:
We denote by $\C_{i}^{(0)}$ the configurations
of size in $[0,\ell^{4})$ and
for $k\ge 1$ we write $\C_{i}^{(k)}$ for the configurations
of size in $[\ell^{k+3},\ell^{k+4})$.
Moreover, define
$\C^{(k)}=\bigcup_i \C_{i}^{(k)}$
and $\C^{(\ge k)} = \bigcup_{h\ge k} \C^{(h)}$.
Let $d$ be the
smallest number such that
$\C^{(\ge d)}$ is empty. Note that
$d\le \log(n) / \log(\ell)$.

Now consider the following random process.
\begin{experiment}\label{exp:sequence}
We construct a nested sequence of resource sets $\I=\I_0 \supseteq \I_1 \supseteq  \ldots \supseteq  \I_d$ as follows. Each $\I_k$ is obtained from $\I_{k-1}$ by deleting every resource in $\I_{k-1}$ independently with probability $(\ell-1) / \ell$.
\end{experiment}
In expectation only a $1/\ell$ fraction of resources in $\I_{k-1}$ survives in $\I_k$.
Also notice that for $C \in \C^{(k)}$ we
have that $\mathbb E[ |\I_k \cap C| ] = \mathrm{poly}(\ell)$.

The proof of Theorem~\ref{thm:unweighted_hypergraph} is organized as follows.
In Section~\ref{sec:sequence}, we give some properties
of the resource sets constructed by
Random Experiment~\ref{exp:sequence} that hold
with high probability.
Then in Section~\ref{sec:LLL}, we show that we can
find a single consistent set of configurations for each group of players such
that for each configuration selected, its intersection with smaller selected configurations
is bounded if we restrict the resource set to an appropriate
$\I_k$. Restricting the resource set is important to bound
the dependencies of bad events in order to apply Lovasz
Local Lemma.
Finally in Section~\ref{sec:reconstruction},
we demonstrate how these configurations allows us to reconstruct
a consistent $\alpha$-relaxed perfect matching for an
appropriate assignment of resources to configurations.

\subsection{Properties of resource sets}\label{sec:sequence}
In this subsection, we give a precise statement of the key properties that we need from Random Experiment~\ref{exp:sequence}. The first two lemmas have a straight-forward proof. The last one is a generalization of an argument used by Bansal and Srividenko \cite{BansalSrividenko}. Since the proof is more technical and tedious, we also defer it to Appendix~\ref{appendix_sequence} along with the proof of the first two statements.

We start with the first property which bounds the size of the 
configurations when restricted to some $\I_k$. This property is useful to reduce the dependencies while applying LLL later. 
\begin{lemma}
\label{lma-size}
Consider Random Experiment~\ref{exp:sequence}
with $\ell\geq 300.000\log^{3} (n)$.
For any $k\geq 0$ and any $C\in\C^{(\geq k)}$ we have 
  \begin{equation*}
      \frac{1}{2} \ell^{-k}|C| \le |\I_k \cap C| \le \frac{3}{2} \ell^{-k}|C|
  \end{equation*}
with probability at least $1-1/n^{10}$.
\end{lemma}
The next property expresses that for any configuration the sum of intersections with configurations of a particular size does not deviate much from its expectation. In particular, for any configuration $C$, the sum of it's intersections with other configurations is at most $|C|\ell$ as each resource is in atmost $\ell$ configurations. By the lemma stated below, we recover this up to a multiplicative constant factor when we consider the appropriately weighted sum of the intersection of $C$ with other configurations $C'$ of smaller sizes where each configuration $C' \in \C^{(k)}$ is restricted to the resource set $\I_k$.

\begin{lemma}
\label{lma-overlap-representative}
Consider Random Experiment~\ref{exp:sequence}
with $\ell\geq 300.000\log^{3} (n)$.
For any $k\geq 0$ and any $C\in\C^{(\geq k)}$ we have 
  \begin{equation*}
      \sum_{C'\in \C^{(k)}} |C'\cap C\cap \I_k| \leq \frac{10}{\ell^{k}} \left(|C|+\sum_{C'\in \C^{(k)}} |C'\cap C| \right)
  \end{equation*}
with probability at least $1-1/n^{10}$.
\end{lemma}
We now define the notion of \emph{good} solutions which is helpful in stating our last property.
Let $\F$ be a set of configurations,
$\alpha:\F \rightarrow \mathbb N$, $\gamma \in\mathbb N$, and $\I'\subseteq \I$.
We say that an assignment of $\I'$ to $\F$ is $(\alpha,\gamma)$-good if every configuration $C\in \F$ receives at least $\alpha(C)$ resources of $C\cap \I'$ and if no resource in $\I'$ is assigned more than $\gamma$ times in total.

Below we obtain that given a $(\alpha,\gamma)$-good solution with respect to resource set $\I_{k+1}$, one can construct an almost $(\ell \cdot \alpha,\gamma)$-good solution with respect to the bigger resource set $\I_{k}$. Informally, starting from a good solution with respect to the final resource set and iteratively applying this lemma would give us a good solution with respect to our complete set of resources.   
\begin{lemma}
\label{lma-good-solution}
Consider Random Experiment~\ref{exp:sequence}
with $\ell\geq 300.000\log^{3} (n)$. Fix $k\geq 0$. Conditioned on the event that the bounds in Lemma~\ref{lma-size} hold
for $k$, then with probability at least $1 - 1/n^{10}$
the following holds for all
$\F\subseteq \C^{(\geq k+1)}$, $\alpha:\F \rightarrow \mathbb N$, and $\gamma \in\mathbb N$ such that $\ell^3/1000\leq \alpha(C) \leq n $ for all $C\in\F$ and
$\gamma \in \{1,\dotsc,\ell\}$:
If there is a $(\alpha,\gamma)$-good assignment of $\I_{k+1}$ to $\F$, then there is a $(\alpha',\gamma)$-good assignment of $\I_k$ to $\F$ where
\begin{equation*}
    \alpha'(C) \ge \ell \left(1-\frac{1}{\log (n)} \right) \alpha(C)
\end{equation*}
for all $C\in\F$.
Moreover, this assignment can be found in polynomial time.
\end{lemma}

Given the lemmata above, by a simple union bound one gets that all the properties of resource sets hold.

\subsection{Selection of configurations}\label{sec:LLL}
In this subsection, we give a random process that selects one consistent set of configurations
for each group of players such that the intersection of the selected configurations with smaller configurations is bounded when
considered on appropriate sets $R_k$. We will denote $\s_A$ the selected consistent set for group $A$ and for ease of notation we will denote $K_i=\s_A^{(i)}$ the selected configuration for player $i\in A$. For any integer $k$, we write $\mathcal K^{(k)}_i = \{K_i\}$ if $K_i\in\mathcal C^{(k)}_i$ and $\mathcal K^{(k)}_i = \emptyset$ otherwise. As for the configuration set, we will also denote $\mathcal K^{(k)}=\bigcup_{i}\mathcal K^{(k)}_i$ and $\mathcal K= \bigcup_{k}\mathcal K^{(k)}$.
The following lemma describes what are the properties we want to have while selecting the configurations. For better clarity we also recall what the properties of the sets
$R_0,\dotsc,R_d$ that we need are. These hold with high probability by the lemmata of
the previous section.

\begin{lemma}\label{lma:main-LLL}
  Let $\I =\I_0\supseteq\dotsc\supseteq\I_d$ be
  sets of fewer and fewer resources.
  Assume that for each $k$ and $C\in \mathcal C_i^{(k)}$ 
  we have
  \begin{equation*}
      1/2 \cdot \ell^{k - h} \le |C\cap \I_h| \le 3/2 \cdot \ell^{- h} |C| < 3/2 \cdot \ell^{k - h + 4}
  \end{equation*}
  for all $h=0,\dotsc,k$.
  Then there exists a selection of one consistent set $\s_A$ for each group $A$ such for all $k=0,\dotsc, d$, $C\in \mathcal C^{(k)}$ and $j=0,\dotsc,k$ then we have
  \begin{equation*}
      \sum_{j\leq h\le k} \sum_{K\in\mathcal K^{(h)}} \ell^{h} |K \cap C \cap \I_h|
      \le \frac{1}{\ell} \sum_{j\leq h\le k} \sum_{C'\in\mathcal C^{(h)}} \ell^{h} |C' \cap C \cap \I_h| + 1000 \frac{d + \ell}{\ell}\log(\ell) |C| .
  \end{equation*}
  Moreover, this selection of consistent sets can be found in polynomial time.
\end{lemma}
Before we prove this lemma, we give an intuition of the statement.
Consider the sets $\I_1,\dotsc,\I_d$ constructed as in Random Experiment~\ref{exp:sequence}.
Then for $C'\in\mathcal C^{(h)}$ we have $\E[\ell^h |C'\cap C\cap \I_h|] = |C'\cap C|$.
Hence
\begin{equation*}
  \sum_{h\le k} \sum_{K\in\mathcal K^{(h)}} |K \cap C| = \E[\sum_{h\le k} \sum_{K\in\mathcal K^{(h)}} \ell^h |K \cap C \cap \I_h|]
\end{equation*}
Similarly for the right-hand side we have
\begin{multline*}
  \E[\frac{1}{\ell} \sum_{j \le h\le k} \sum_{C'\in\mathcal C^{(h)}} \ell^h |C' \cap C \cap \I_h| + O(\frac{d + \ell}{\ell}\log(\ell) |C|)] \\
  = \frac{1}{\ell}\underbrace{\sum_{j\le h\le k} \sum_{C'\in\mathcal C^{(h)}} |C' \cap C|}_{\le \ell |C|} + O\left(\frac{d + \ell}{\ell}\log(\ell) |C|\right)
  = O\left(\frac{d + \ell}{\ell}\log(\ell) |C|\right) .
\end{multline*}
Hence the lemma says that each resource in $C$ is roughly covered $O((d + \ell)/\ell \cdot \log(\ell))$ times by smaller configurations.

We now proceed to prove the lemma by performing the following random experiment
and by Lovasz Local Lemma show that there
is a positive probability of success.
\begin{experiment}
For each group $A$, select one consistent set $\s_A$ uniformly at random. Then for each player $i \in A$ set $K_i=\mathcal \s_A^{(i)}$.
\end{experiment}
For all $h=0,\dotsc,d$ and $i\in P$
we define the random variable
\begin{equation*}
    X^{(h)}_{i,C} = \sum_{K\in\mathcal K^{(h)}_i} |K \cap C \cap \I_h| \le \min\{3/2 \cdot \ell^4, |C\cap \I_h|\} .
\end{equation*}
Let $X^{(h)}_C = \sum_{i=1}^m X^{(h)}_{i, C}$.
Then
\begin{equation*}
  \E[X^{(h)}_C] \le \frac{1}{\ell} \sum_{C'\in\mathcal C^{(h)}} |C'\cap C\cap \I_h| \le |C\cap \I_h| .
\end{equation*}
We define a set of bad events. As we will show
later, if none of them
occur, the properties from the premise hold.
For each $k$, $C\in\mathcal C^{(k)}$, and $h\le k$ let $B_C^{(h)}$ be the event that
\begin{equation*}
    X_C^{(h)}
    \ge \begin{cases}
      \E[X_C^{(h)}] + 63 |C\cap \I_h| \log(\ell) &\text{ if $k - 5 \le h \le k$}, \\
      \E[X_C^{(h)}] + 135 |C\cap \I_h| \log(\ell) \cdot \ell^{-1} &\text{ if $h \le k - 6$}.
    \end{cases}
\end{equation*}
There is an intuitive reason as to why we define these two different bad events. In the case $h\leq k-6$, we are counting how many times $C$ is intersected by configurations that are much smaller than $C$. Hence the size of this intersection can be written as a sum of independent random variables of value at most $O(\ell^4)$ which is much smaller than the total size of the configuration $|C\cap \I_h|$. Since the random variables are in a much smaller range, Chernoff bounds give much better concentration guarantees and we can afford a very small deviation from the expectation. In the other case, we do not have this property hence we need a bigger deviation to maintain a sufficiently low probability of failure. However, this does not hurt the statement of Lemma~\ref{lma:main-LLL} since we sum this bigger deviation only a constant number of times. With this intuition in mind, we claim the following.
\begin{claim}
For each $k$, $C\in\mathcal C^{(k)}$, and $h\le k$ we have
\begin{equation*}
    \P[B_C^{(h)}] \le \exp\left(- 2\frac{|C \cap \I_h|}{\ell^9} - 18\log(\ell)\right) .
\end{equation*}
\end{claim}

\begin{proof}
Consider first the case that $h \ge k - 5$.
By a Chernoff bound (see Proposition~\ref{cor:chernoff}) with
\begin{equation*}
    \delta = 63\frac{|C\cap \I_h| \log(\ell)}{\E[X_C^{(h)}]} \ge 1
\end{equation*}
we get
\begin{equation*}
    \P[B_C^{(h)}] \le \exp\bigg(-\frac{\delta \E[X^{(h)}_C]}{3 |C\cap \I_h|}\bigg) \le \exp(-21\log(\ell))) \le \exp\bigg(-2 \underbrace{\frac{|C\cap \I_h|}{\ell^{9}}}_{\le 3/2} - 18\log(\ell)\bigg).
\end{equation*}
Now consider $h \le k - 6$.
We apply again a Chernoff bound with
\begin{equation*}
    \delta = 135\frac{|C\cap\I_h| \log(\ell)}{\ell \E[X_C^{(h)}]} \ge \frac{1}{\ell} .
\end{equation*}
This implies
\begin{multline*}
    \mathbb P[B_C^{(h)}]
    \le \exp\left(-\frac{\min\{\delta,\delta^2\} \E[X^{(h)}_C]}{3 \cdot 3/2 \cdot \ell^4}\right)
    \le \exp\left(-30\frac{|C\cap\I_h| \log(\ell)}{\ell^6} \right) \\
    \le \exp\left(-2 \frac{|C\cap \I_h|}{\ell^9} - 18\log(\ell)\right) . \qedhere
\end{multline*}
\end{proof}
\begin{proposition}[Lovasz Local Lemma (LLL)]\label{prop:LLL}
Let $B_1, \dotsc, B_t$ be bad events, and let
$G = (\{B_1,\dotsc,B_t\}, E)$ be a dependency graph for them, in which for every $i$, event $B_i$ is mutually independent of all events $B_j$ for which $(B_i, B_j)\notin E$.
Let $x_i$ for $1\le i \le t$ be such that
$0 < x(B_i) < 1$ and
$\P[B_i]\le x(B_i) \prod_{(B_i,B_j)\in E} (1-x(B_j))$.
Then with positive probability no event $B_i$ holds.
\end{proposition}
Let $k\in\{0,\dotsc,d\}$, $C\in\mathcal C^{(k)}$ and $h\le k$.
For event $B_C^{(h)}$ we set
\begin{equation*}
    x(B_C^{(h)}) = \exp(-|C\cap\I_h| / \ell^9 - 18\log(\ell)) .
\end{equation*}
We now analyze the dependencies of $B_C^{(h)}$.
The event depends only on random variables $\s_A$ for groups $A$ that contain at least one player $i$ that has a configuration in $\mathcal C^{(h)}_i$ which overlaps with $C\cap \I_h$.
The number of such configurations (in particular,
of such groups) is at most
$\ell |C\cap \I_h|$ since the hypergraph is regular.

In each of these groups, we count at most $\log (n)$ players, each having $\ell$ configurations hence in total at most $\ell\cdot \log (n)$ configurations. 

Each configuration $C'\in \C^{(h')}$ can only influence those events $B^{(h')}_{C''}$ where $C' \cap C'' \cap \I_{h'} \neq \emptyset$. Since $|C'\cap \I_{h'}|\leq 3/2\cdot \ell^4$ and since each resource appears in at most $\ell$ configurations, we see that each configuration can influence at most $3/2 \cdot \ell^5$ events.

Putting everything together, we see that the bad event $B_C^{(h)}$ is independent of all but at most
\begin{equation*}
    (\ell |C\cap \I_h|) \cdot (\ell\cdot \log (n)) \cdot (3/2 \cdot \ell^5) = 3/2\cdot  \ell^7 \cdot \log (n) |C\cap \I_h|  \leq |C\cap \I_h|\ell^8
\end{equation*} other bad events.

We can now verify the condition for Proposition~\ref{prop:LLL} by calculating
\begin{align*}
    x(B_C^{(h)}) & \prod_{(B_C^{(h)}, B_{C'}^{(h')})\in E} (1 - x(B_{C'}^{(h')})) \\
    &\ge \exp(-|C\cap\I_h|/\ell^9 - 18\log(\ell)) \cdot (1 - \ell^{-18})^{|C\cap\I_h|\ell^8} \\
    &\ge \exp(-|C\cap\I_h|/\ell^9 - 18\log(\ell)) \cdot \exp(- |C\cap\I_h| / \ell^9) \\
    &\ge \exp(-2|C\cap\I_h|/\ell^9 - 18\log(\ell)) \ge \P[B^{(h)}_C] .
\end{align*}
By LLL we have that with positive probability none
of the bad events happen. Let $k\in\{0,\dotsc,d\}$ and $C\in\mathcal C^{(k)}$.
Then for $k - 5 \le h \le k$ we have
\begin{equation*}
  \ell^{h} X^{(h)}_C \le \ell^{h} \E[X_C^{(h)}] + 63 \ell^{h} |C\cap\I_h|\log(\ell)
    \le \ell^{h} \E[X_C^{(h)}] + 95 |C|\log(\ell) .
\end{equation*}
Moreover, for $h\le k-6$ it holds that
\begin{equation*}
  \ell^{h} X^{(h)}_C \le \ell^{h} \E[X_C^{(h)}] + 135 \ell^{h-1} |C\cap\I_h|\log(\ell)
    \le \ell^{h} \E[X_C^{(h)}] + 203 |C|\log(\ell) \cdot \ell^{-1} .
\end{equation*}
We conclude that, for any $0\leq j\leq k$,
\begin{align*}
    \sum_{j\leq h\le k} \sum_{K\in\mathcal K^{(h)}}
 \ell^{h} |K \cap C \cap \I_h|
 &\le \sum_{j\leq h\le k} \ell^{h}  \E[X^{(h)}_{C}] + 1000 \frac{(k-j + 1) + \ell}{\ell} |C| \log(\ell) \\
 &\le \frac{1}{\ell} \sum_{j\leq h\le k} \ell^{h} \sum_{C'\in\mathcal C^{(h)}} |C'\cap C \cap \I_h| + 1000 \frac{d + \ell}{\ell} |C| \log(\ell) .
\end{align*}
This proves Lemma~\ref{lma:main-LLL}.

\begin{remark}{\rm
Since there are at most $\mathrm{poly}(n,m,\ell)$ bad events and each bad event $B$ has $\frac{x(B)}{1-x(B)}\le1/2$ (because $x(B)\le \ell^{-18}$), the constructive
variant of LLL by Moser and Tardos~\cite{moser2010constructive} can be applied to find a selection of configurations such that no bad events occur in randomized polynomial time.}
\end{remark}

\subsection{Assignment of resources to configurations}\label{sec:reconstruction}
In this subsection, we show how all the previously established properties allow us to find, in polynomial time, a good assignment of resources to the configurations $\K$ chosen as in the previous subsection. We will denote as in the previous subsection $\K_i^{(k)}=\{K_i\}$ if $K_i\in \C_i^{(k)}$ and $\K_i^{(k)}=\emptyset$ otherwise. We also define $\K^{(k)}=\bigcup_{i}\K_i^{(k)}$ and $\K^{(\geq k)}=\bigcup_{h\geq k}\K^{(k)}$.
Finally we define the parameter
\begin{equation*}
    \gamma = 100.000 \frac{d+\ell}{\ell}\log(\ell) ,
\end{equation*}
which will define how many times each resource can be assigned to configurations in an intermediate solution. Note that $d\le\log(n)/\log(\ell)$. By our choice of $\ell=300.000\log^3(n)$, we have that $\gamma \leq 310.000 \log \log (n)$.
Lemma~\ref{lma:main-LLL} implies the following
bound. For sake of brevity, the proof is deferred to Appendix~\ref{appendix_reconstruct}.
\begin{claim}
\label{cla:reconstruct}
For any $k\geq 0$, any $0\leq j\leq k$, and any $C\in \K^{(k)}$
\begin{equation*}
    \sum_{j\leq h\leq k}\sum_{K\in \K^{(h)}} \ell^{h}|K\cap C \cap \I_h| \leq 2000\frac{d+\ell}{\ell}\log (\ell) |C|
\end{equation*}
\end{claim}

The main technical part of this section is
the following lemma that is proved by induction.
\begin{lemma}
\label{lem:reconstruct}
For any $j\geq 0$, there exists an assignment of resources of $\I_j$ to configurations in $\K^{(\geq j)}$ such that no resource is taken more than $\gamma$ times and each configuration $C\in \K^{(k)}$ ($k\geq j$) receives at least
\begin{equation*}
     \left(1-\frac{1}{\log (n)} \right)^{2(k-j)}\ell^{k-j} |C\cap \I_k|-\frac{3}{\gamma}\sum_{j\leq h\leq k} \sum_{K\in \K^{(h)}} \ell^{h-j}|K\cap C \cap \I_h|
\end{equation*}
resources from $\I_k$.
\end{lemma}

Before proceeding to the proof, we first give intuition of why this is what we want to prove. Note that the term $\ell^{k-j}|C\cap \I_k|$ is roughly equal to $\ell^{-j}|C|$ by the properties of the resource sets (precisely Lemma \ref{lma-size}). The second term 
\begin{equation*}
    \sum_{j\leq h\leq k} \sum_{K\in \K^{(h)}} \ell^{h-j}|K\cap C \cap \I_h|
\end{equation*}
can be shown to be 
\begin{equation*}
    O\left(\ell^{-j}\frac{d+\ell}{\ell}\log (\ell) |C| \right)= O (\ell^{-j}\log \log (n) |C|)
\end{equation*}
by Claim \ref{cla:reconstruct}. Hence by choosing $\gamma$ to be $\Theta (\log \log (n))$ we get that the bound in Lemma \ref{lem:reconstruct} will be $\Theta (\ell^{-j}|C|)$. At the end of the induction, we have $j=0$ which indeed implies that we have an assignment in which configurations receive
\begin{equation*}
    \Theta (\ell^{-0}|C|)=\Theta(|C|)
\end{equation*} resources and such that each resource is assigned to at most $O (\log \log (n))$ configurations.

\begin{proof}
We start from the biggest configurations and then iteratively reconstruct a good solution for smaller and smaller configurations. Recall $d$ is the smallest integer such that $\K^{(\geq d)}$ is empty.
Our base case for these configurations in $\K^{(\geq d)}$ is vacuously satisfied.

Now assume that we have a solution at level $j$, i.e. an assignment of resources to configurations in $\K^{(\geq j)}$ such that no resource is taken more than $\gamma$ times and each configuration $C\in \K^{(k)}$ such that $k\geq j$ receives at least
\begin{equation*}
     \left(1-\frac{1}{\log (n)} \right)^{2(k-j)}\ell^{k-j} |C\cap \I_k|-\frac{3}{\gamma}\sum_{j\leq h\leq k} \sum_{K\in \K^{(h)}} \ell^{h-j}|K\cap C \cap \I_h|
\end{equation*}
resources from $\I_j$.
We show that this implies a solution at level $j-1$ in the following way. First by Lemma~\ref{lma-good-solution}, this implies an assignment of resources of $\I_{j-1}$ to configurations in $\K^{(\geq j)}$ such that each $C\in \K^{(k)}$ receives at least
\begin{align*}
     &\left(1-\frac{1}{\log (n)} \right)\ell \left(\ell^{k-j} \left(1-\frac{1}{\log (n)} \right)^{2(k-j)} |C\cap \I_k|-\frac{3}{\gamma}\sum_{j\leq h\leq k} \sum_{K\in \K^{(h)}} \ell^{h-j}|K\cap C \cap \I_h|\right)\\
     &=\left(1-\frac{1}{\log (n)} \right)^{2(k-(j-1))-1} \ell^{k-(j-1)} |C\cap \I_k|-\frac{3}{\gamma}\left(1-\frac{1}{\log (n)} \right)\sum_{j\leq h\leq k} \sum_{K\in \K^{(h)}} \ell^{h-(j-1)}|K\cap C \cap \I_h|\\
     &\geq  \left(1-\frac{1}{\log (n)} \right)^{2(k-(j-1))-1}\ell^{k-(j-1)} |C\cap \I_k|-\frac{3}{\gamma}\sum_{j\leq h\leq k} \sum_{K\in \K^{(h)}} \ell^{h-(j-1)}|K\cap C \cap \I_h|
\end{align*}
resources and no resource of $\I_{j-1}$ is taken more than $\gamma$ times. Note that we can apply Lemma \ref{lma-good-solution} since we have by Claim \ref{cla:reconstruct} and Lemma \ref{lma-size}
\begin{align*}
        &\left(1-\frac{1}{\log (n)} \right)^{2(k-j)}\ell^{k-j} |C\cap \I_k|-\frac{3}{\gamma}\sum_{j\leq h\leq k} \sum_{K\in \K^{(h)}} \ell^{h-j}|K\cap C \cap \I_h| \\
        &\geq \frac{\ell^{k-j}}{e^2}|C\cap R_k| - \frac{3}{\gamma}2000\ell^{-j}\frac{d+\ell}{\ell}\log(\ell)|C|\\
        &\geq \ell^{-j}|C|\left(\frac{1}{2e^2}-\frac{6000}{\gamma}\frac{d+\ell}{\ell}\log(\ell)\right)\\
        &\geq \frac{\ell^{-j}|C|}{3e^2}>\frac{\ell^3}{1000}
\end{align*}
Now consider configurations in $\K^{(j-1)}$ and proceed for them as follows. Give to each $C\in\K^{(j-1)}$ all the resources in $C\cap \I_{j-1}$ except all the resources that appear in more than $\gamma$ configurations in $\K^{(j-1)}$. Since each deleted resource is counted at least $\gamma$ times in the sum $\sum_{K\in \K^{(j-1)}}|K\cap C\cap \I_{j-1}|$, we have that each configuration $C$ in $\K^{(j-1)}$ receives at least 
\begin{equation*}
    |C\cap \I_{j-1}|-\frac{1}{\gamma}\sum_{K\in \K^{(j-1)}}|K\cap C\cap \I_{j-1}|
\end{equation*}
resources and no resource is taken more than $\gamma$ times by configurations in $\K^{(j-1)}$. Notice that now every resource is taken no more than $\gamma$ times by configurations in $\K^{(\geq j)}$ and no more than $\gamma$ times by configurations in $\K^{(j-1)}$ which in total can sum up to $2\gamma$ times. 

Therefore to finish the proof consider an resource $i\in \I_{j-1}$. This resource is taken $b_i$ times by configurations in $\K^{(\geq j)}$ and $a_i$ times by configurations in $\K^{(j-1)}$. If $a_i+b_i \leq \gamma$, nothing needs to be done. Otherwise, denote by $O$ the set of problematic resources (i.e. resources $i$ such that $a_i+b_i>\gamma$). For every $i\in O$, select uniformly at random $a_i+b_i-\gamma$ configurations in $\K^{(\geq j)}$ that currently contain resource $i$ and delete the resource from these configurations. When this happens, each configuration in $C\in \K^{(\geq j)}$ that contains $i$ has a probability of $(a_i+b_i-\gamma)/b_i$ to be selected to loose this resource. Hence the expected number of resources that $C$ looses with such a process is

\begin{equation*}
    \mu = \sum_{i\in O\cap C} \frac{a_i+b_i-\gamma}{b_i}
\end{equation*}
It is not difficult to prove the following claim. However, for better clarity we defer its proof to appendix \ref{appendix_reconstruct}.
\begin{claim} For any $C\in \K^{(\geq j)}$,
\label{cla:reconstruct_mu}
\begin{equation*}
    \frac{1}{\gamma^2}\sum_{K\in \K^{(j-1)}}|K\cap C \cap \I_{j-1}\cap O|\leq \mu \leq \frac{2}{\gamma} \sum_{K\in \K^{(j-1)}}|K\cap C \cap \I_{j-1}\cap O|
\end{equation*}
\end{claim}
Assume then that $\mu \leq \frac{|C\cap \I_k|}{10^{12}\log^3 (n)}$. Note that $C$ cannot loose more than $\sum_{K\in \K^{(j-1)}}|K\cap C \cap \I_{j-1}\cap O|$ resources in any case. Therefore, by assumption on $\mu$, and since 
\begin{equation*}
    \mu\geq \frac{1}{\gamma^2}\sum_{K\in \K^{(j-1)}}|K\cap C \cap \I_{j-1}\cap O|\ ,
\end{equation*}
we have that 
\begin{equation*}
    \sum_{K\in \K^{(j-1)}}|K\cap C \cap \I_{j-1}\cap O|\leq \frac{\gamma^2}{10^{12}\log^3 (n)} |C\cap \I_k|\leq \frac{10^{11} \log^2 \log (n)}{10^{12}\log^3 (n)}|C\cap \I_k|\leq \frac{1}{\log (n)}|C\cap \I_k|\ .
\end{equation*}
Therefore $C$ looses at most $|C\cap \I_k|/\log (n)$ resources. Otherwise we have that 
\begin{equation*}
    \mu > \frac{|C\cap \I_k|}{10^{12}\log^2 (n)} \geq \frac{\ell^3}{10^{12} \log^3 (n)} \geq 200\log(n)
\end{equation*}
by Lemma~\ref{lma-size}. Hence noting $X$ the number of deleted resources in $C$ we have that 
\begin{equation*}
    \mathbb P\left(X\geq \frac{3}{2}\mu \right) \leq \exp\left(-\frac{\mu}{12} \right)\leq \frac{1}{n^{10}}.
\end{equation*}
With high probability no configuration looses more than 
\begin{equation*}
    \frac{3}{2}\mu \leq \frac{3}{\gamma}\sum_{K\in \K^{(j-1)}}|K\cap C \cap \I_{j-1}\cap O|\leq \frac{3}{\gamma}\sum_{K\in \K^{(j-1)}}|K\cap C \cap \I_{j-1}|
\end{equation*}
resources. Hence each configuration $C\in \K^{(\geq j)}$ ends with at least
\begin{align*}
    &\left(1-\frac{1}{\log (n)} \right)^{2(k-(j-1))-1}\ell^{k-(j-1)} |C\cap \I_k|-\frac{3}{\gamma}\sum_{j\leq h\leq k} \sum_{K\in \K^{(h)}} \ell^{h-(j-1)}|K\cap C \cap \I_h|\\
    &-\frac{1}{\log (n)}\left(1-\frac{1}{\log (n)} \right)^{2(k-(j-1))-1}\ell^{k-(j-1)} |C\cap \I_k| - \frac{3}{\gamma}\sum_{K\in \K^{(j-1)}}|K\cap C \cap \I_{j-1}|\\
    &\geq \left(1-\frac{1}{\log (n)} \right)^{2(k-(j-1))}\ell^{k-(j-1)} |C\cap \I_k|-\frac{3}{\gamma}\sum_{j-1\leq h\leq k} \sum_{K\in \K^{(h)}} \ell^{h-(j-1)}|K\cap C \cap \I_h|
\end{align*}
resources which concludes the proof.
\end{proof}
\begin{corollary}
\label{reconstruct_corollary}
There exists an assignment of resources $\I$ to $\K$ such that each configuration $C\in  \K$ receives at least $\left\lfloor |C|/(100\gamma) \right\rfloor$ resources. Moreover, this assignment can be found in polynomial time.
\end{corollary}
\begin{proof}
Lemma \ref{lem:reconstruct} for $k=0$ and Claim \ref{cla:reconstruct} together imply that
we can assign at least
\begin{equation*}
    \frac{|C|}{2e^2}-\frac{6000}{100.000}|C|\geq \frac{|C|}{100}
\end{equation*}
resources to every $C\in \K$ such that no resource in $\I$ is assigned more than $\gamma$ times. In particular, we can fractionally assign
at least $|C| / (100\gamma)$ resources
to each $C\in \K$ such that no resource
is assigned more than once.
By integrality of the bipartite matching polytope, the corollary follows.
\end{proof}

\section{Further connections between hypergraph matching and Santa Claus}
\label{sec:reduction santa claus}
In Section~\ref{sec:hypergraph problem} we essentially prove that every regular (non-uniform) hypergraph has an $\alpha$-relaxed perfect matching for some $\alpha=O(\log \log (n))$, assuming that all hyperedges contain at least $\alpha$ resources. This means that we give a sufficient condition for a hypergraph to have a good relaxed matching. A natural optimization problem that arises from this is the following:
Given any unweighted hypergraph, which is not necessarily regular nor
all hyperedges necessarily contain many resources, what is the minimum $\alpha$ such that there exists an $\alpha$-relaxed perfect matching in this hypergraph?

In this section, we investigate the relationship between this problem and the Santa Claus problem with linear utility functions. Formally, the two problems considered are precisely the following.

\paragraph{Matching in general hypergraphs.} Consider a (non-uniform) hypergraph $\mathcal H=(P\cup R, \C)$ with unit weights, that is, $w_{j,C}=1$ for all $j,C$ such that $j\in C$. The problem is to find the minimum $\alpha$ such that $\mathcal H$ has an $\alpha$-relaxed perfect matching (and output such a matching).

\paragraph{The Santa Claus with linear utility functions.} In this case, each player $i$ has an arbitrary linear utility function $f_i$. We note that there is no relationship assumed between the utility functions of different players. The goal is to assign resources to players to maximize the minimum utility among players. As mentioned in introduction, the best approximation algorithm for this problem is an $O(n^\epsilon)$-approximation running in time $O(n^{1/\epsilon})$. 
\bigskip

We show by a straightforward reduction that a $c$-approximation for the Santa Claus problem immediately implies a $c$-approximation for the matching problem. Interestingly, there is also a close connection in the opposite direction.
\begin{theorem}\label{thm:reduction}
A $c$-approximation algorithm to the hypergraph matching problem in general hypergraphs yields an
$O((c\log^* (n))^2)$-approximation algorithm to the Santa Claus problem.
\end{theorem}

We mention that we implicitly refer to polynomial time algorithms even when not specified. All the proofs of this section are deferred to Appendix \ref{appendix:reduction}. We also mention that Theorem~\ref{thm:reduction} implies that any sub-polynomial approximation to the matching problem would be a significant improvement of the state-of-the-art for Santa Claus with arbitrary linear utility functions.

\paragraph{Remark.} Since hypergraphs considered here might be non-regular and some hyperedges might contain very few resources, our result in Section \ref{sec:hypergraph problem} does not imply any approximation for the optimization problem considered here. Our reduction in this section makes a crucial use of small hyperedges containing only one resource. This shows that handling the small hyperedges is one of the core difficulties in this case.



\section{Conclusion}

We investigated the submodular Santa Claus in the restricted assignment case and gave a $O(\log \log (n))$-approximation for this problem. This represents a significant generalization of the results for the linear case.
The submodularity of the utility function introduced new obstacles compared to the linear case. These difficulties are captured by the fact that we need to solve a new matching problem in non-uniform hypergraphs that generalizes the case of uniform hypergraphs which has been already studied in the context of the restricted Santa Claus problem with a linear utility function. Under the assumption that the hypergraph is regular and all edges are sufficiently large, we proved that there is always a $\alpha$-relaxed perfect matching for $\alpha = O(\log \log (n))$. This result generalizes the work of Bansal and Srividenko~\cite{BansalSrividenko}.
It remains an intriguing question whether one can get $\alpha = O(1)$ as it is possible in the uniform case.
One idea (similar to Feige's proof in the uniform case~\cite{Feige}) would be to view our proof as a sparsification theorem
and to apply it several times.
Given a set of hyperedges such that every player has $\ell$ hyperedges and every resource appears in no more than $\ell$ hyperedges, one would like to select $\textrm{polylog}(\ell)$ hyperedges for each player such that all resources appear in no more than $\textrm{polylog}(\ell)$ of the selected hyperedges. It is not difficult to see than our proof actually achieves this when $\ell=\textrm{polylog}(n)$. However, repeating this after the first step seems to require new ideas since our bound on the number of times each resource is taken is $\Omega \left(\frac{d+\ell}{\ell}\log(\ell) \right)$ where $\ell$ is the current sparsity and $d$ the number of configuration sizes. For the first step, we conveniently have that $d=O(\log (n))=O(\ell)$ but after the first sparsification, it may not be true.

We also provided a reduction from the Santa Claus with arbitrary linear utility functions to the hypergraph matching problem in general hypergraphs. This shows that finding the smallest $\alpha$ such that a hypergraph has an $\alpha$-relaxed perfect matching (or approximating it) is a very non-trivial problem (even within a sub-polynomial factor).
Another interesting question is to improve the $O(\log^*(n))^2$ factor in the reduction to a constant.

\section{Acknowledgements}
The authors wish to thank Ola Svensson for helpful discussions on the problem.

\bibliographystyle{alpha}
\bibliography{refs}

\appendix

\section{Concentration bounds}
\begin{proposition}[Chernoff bounds (see e.g.~\cite{mitzenmacher2017probability})]
\label{chernoff}
Let $X=\sum_i X_i$ be a sum of independent random variables such that each $X_i$ can take values in a range $[0,1]$. Define $\mu=\mathbb E(X)$. We then have the following bounds 

\begin{equation*}
    \mathbb P \left(X\geq (1+\delta)\mathbb E(X) \right) \leq \exp\left(-\frac{\min\{\delta,\delta^2\} \mu}{3} \right)
\end{equation*} for any $\delta>0$.
\begin{equation*}
    \mathbb P \left(X\leq (1-\delta)\mathbb E(X) \right) \leq \exp\left(-\frac{\delta^2 \mu}{2} \right)
\end{equation*} for any $0<\delta<1$.
\end{proposition}

The following proposition follows immediately from Proposition \ref{chernoff} 
by apply it with $X'=X/a$.
\begin{proposition}
\label{cor:chernoff}
Let $X=\sum_i X_i$ be a sum of independent random variables such that each $X_i$ can take values in a range $[0,a]$ for some $a>0$. Define $\mu=\mathbb E(X)$. We then have the following bounds 

\begin{equation*}
    \mathbb P \left(X\geq (1+\delta)\mathbb E(X) \right) \leq \exp\left(-\frac{\min\{\delta,\delta^2\} \mu}{3a} \right)
\end{equation*} for any $\delta>0$.
\begin{equation*}
    \mathbb P \left(X\leq (1-\delta)\mathbb E(X) \right) \leq \exp\left(-\frac{\delta^2 \mu}{2a} \right)
\end{equation*} for any $0<\delta<1$.
\end{proposition}


\section{Omitted proofs from Section~\ref{sec:reduction to hypergraph}}\label{appendix_lp}
\subsection{Solving the configuration LP}
The goal of this section is to prove Theorem~\ref{thm:config-LP}.
We consider the dual of the configuration LP (after adding an artificial minimization direction $\min 0^T x$).
\begin{align*}
    \max \sum_{i\in P} y_i &- \sum_{j\in R} z_j \\
    \sum_{j\in C} z_j &\ge y_i \quad \text{ for all } i\in P, C\in\C(i, T) \\
    y_j, z_i &\ge 0
\end{align*}
Observe that the optimum of the dual is either $0$ obtained
by $y_i = 0$ and $z_j = 0$ for all $i,j$ or it is unbounded:
If it has any solution with $\sum_{i\in P} y_i - \sum_{j\in R} z_j > 0$, the variables can be scaled by an arbitrary common factor
to obtain any objective value.
If it is unbounded, this can therefore
be certified by providing a feasible solution $y, z$ with
\begin{equation}
    \sum_{i\in P} y_i - \sum_{j\in R} z_j \ge 1 \tag{$*$}.
\end{equation}
We approximate the dual in the variant with constraint $(*)$
instead of a maximization direction using the ellipsoid method.
The separation problem of the dual is as follows.
Given $z_j$, $y_i$ find a player $i$ and set $C$ with
$g(C\cap \Gamma_i) \ge T$ such that $\sum_{j\in C} z_j < y_i$.

To this end,
consider the related problem of maximizing a monotone submodular
function subject to knapsack constraints.
In this problem we are given a monotone submodular function $g$ over
a ground set $E$ and the goal is to maximize $g(E')$ over
all $E'\subseteq E$ with $\sum_{j\in E'} a_j \le b$.
Here $a_j \ge 0$ is a weight associated with $j\in E$ and $b$ is a capacity.
For this problem Srividenko gave a polynomial time
$(1 - 1/e)$-approximation algorithm~\cite{DBLP:journals/orl/Sviridenko04}.
It is not hard to see that this can be used to give a constant
approximation for the variation where strict inequality is required
in the knapsack constraint: Assume w.l.o.g.\ that $0 < a_j < b$ for
all $j$. Then run Srivideko's algorithm to find a set $E'$
with $\sum_{j\in E'} a_j \le b$. Notice that
$g(E')$ is at least $(1 - 1/e)\OPT$, also when $\OPT$ is the optimal
value with respect to strict inequality.
If $E'$ contains only one element
then equality in the knapsack constraint cannot hold and we are done.
Otherwise, split $E'$ into two arbitrary
non-empty parts $E''$ and $E'''$. It follows that
$\sum_{j\in E''} a_j < b$ and $\sum_{j\in E'''} a_j < b$.
Moreover, either $g(E'') \ge g(E') / 2$ or $g(E'') \ge g(E') / 2$.
Hence, this method yields a $c$-approximation for $c = (1 - 1/e)/2$.
We now demonstrate how to use this to find a $c$-approximation
to the configuration LP.

Let $\OPT$ be the optimum of the configuration LP.
It suffices to solve the problem of finding for a given $T$
either a solution of value $c T$ or deciding that $T > \OPT$.
This can then be embedded into a standard dual approximation framework.
We run the ellipsoid method on the dual of the configuration LP
with objective value $c T$ and constraint $(*)$.
This means we have to solve the separation
problem.
Let $z, y$ be the variables at some state.
We first check whether $(*)$ is satisfied, that
is $\sum_{i\in P} y_i - \sum_{j\in R} z_j \ge 1$.
If not, we return this inequality as a separating hyperplane.
Hence, assume $(*)$ is satisfied and our goal
is to find a violated constraint
of the form $\sum_{j\in C} z_j < y_i$ for some
$i\in P$ and $C\in \C(i, T)$.
For each player $i$ we maximize $f$ over all $S\subseteq \Gamma_i$
with $\sum_{j\in S} z_j < y_i$. We use the variant of
Srividenko's algorithm described above to obtain a $c$-approximation
for each player. If for one player $i$ the resulting set $S$ satisfies
$f(S) \ge c T$, then we have found a separating hyperplane to provide
to the ellipsoid method. Otherwise, we know that $f(S) < T$
for all players $i$ and $S\subseteq\Gamma_i$ with
$\sum_{j\in S} z_j < y_i$. In other words, for all players $i$
and all $C\in\C(i,T)$ it holds that $\sum_{j\in C} z_j \ge y_i$,
i.e., $z, y$ is feasible for objective value $T$ and hence
$\OPT > T$. If the ellipsoid method terminates without concluding
that $\OPT > T$, we can derive a feasible primal solution
with objective value $cT$:
The configurations constructed for separating
hyperplanes suffice to prove that the dual is bounded.
These configurations can only be polynomially many by the polynomial
running time of the ellipsoid method.
Hence, when restricting the primal to these configurations
it must remain feasible. To obtain the primal solution we now
only need to solve a polynomial size linear program.
This concludes the proof of Theorem~\ref{thm:config-LP}.

\subsection{Clusters}
This section is devoted to prove Lemma~\ref{lem:config-sample}.
The arguments are similar to those used in~\cite{BansalSrividenko}.

\begin{lemma}
\label{lem:clusters} 
Let $x^*$ be a solution to the configuration LP
of value $T^*$. Then $x^*$ can be transformed into some $x'_{i,C} \ge 0$
for $i\in P$, $C\in\C_t(i, T^*)$ which satisfies the following. There is a partition of the
players into clusters 
$K_1\cup\cdots \cup K_k \cup Q = P$ that satisfy the following. 
\begin{enumerate}
    \item any thin resource $j$ is fractionally assigned at most once, that is, 
    \begin{equation*}
        \sum_{i\in P} \sum_{C\in \C_t(i,T^*):j\in C}x'_{i, C} \le 1
    \end{equation*} We say that the \textit{congestion} on item $j$ is at most 1.  
    \item every cluster $K_j$ gets at least $1/2$ thin configurations in $x'$, that is,
\begin{equation*}
    \sum_{i\in K_j} \sum_{C\in \C_t(i,T^*)} x'_{i, C} \ge 1/2 ;
\end{equation*}
    \item given any $i_1\in K_1, i_2\in K_2,\dotsc,i_k\in K_k$
    there is a matching of fat resources to players
    $P\setminus\{i_1,\dotsc,i_k\}$ such that each of these players $i$ gets a unique fat resource $j\in\Gamma_i$.
\end{enumerate}
\end{lemma}
The role of the set of players $Q$ in the lemma above is that each of them
gets one fat resource for certain.
\begin{proof}
We first transform the solution $x^*$ as follows. For every configuration $C$ (for player $i$) that contains at least one fat resource and such that $x^*_{i,C}>0$, we select arbitrarily one of these fat resources $j$ and we set $x^*_{i,\{j\}}=x^*_{i,C}$ and then we set $x^*_{i,C}=0$. It is clear that this does not increase the congestion on resources and now every configuration that has non-zero value is either a thin configuration or a singleton containing one fat resource.
Therefore we can consider the bipartite graph $G$ formed between the players and the fat resources where there is an edge between player $i$ and fat resource $j$ if the corresponding configuration $C=\{j\}$ is of non zero value (i.e. $x^*_{i,C}>0$). The value of such an edge will be exactly the value $x^*_{i,C}$.
We now make G acyclic by doing the following operation until there exists no cycle anymore. Pick any cycle (which must have even length since the graph is bipartite) and increase the coordinate of $x^*$ corresponding to every other edge in the cycle by a small constant. Decrease the value corresponding to the remaining edges of the cycle by the same constant. This ensures that fat resources are still (fractionally) taken at most once and that the players still have one unit of configurations fractionally assigned to them. We continue this until one of the edge value becomes 0 or 1. If an edge becomes 0, delete that edge and if it becomes 1, assign the corresponding resource to the corresponding player forever. Then delete the player and the resource from the graph and add the player to the cluster $Q$. By construction, every added player to $Q$ is assigned a unique fat resource. Notice that when we stop, each remaining player still has at least 1 unit of configurations assigned to him and every fat resource is still (fractionally) taken at most once. Hence we get a new assignment vector where the assignments of fat resources to players form a forest. We also note that the congestion on thin resources did not increase during this process (it actually only decreased either when we replace fat configurations by a singleton and when players are put into the set $Q$ and deleted from the instance).
We show below how to get the clusters for any tree in the forest.
\begin{enumerate}
\item If the tree consists of a single player, then it trivially forms its own cluster. By feasibility of the original solution $x^*$, condition 2 of the lemma holds. 

\item If there is a fat resource that has degree 1, assign it to its player, add the player to $Q$ and delete both the player and resource. Continue this until every resource has a degree of at least 2. This step adds players to cluster $Q$. By construction, every added player is assigned a unique fat resource. 

\item While there is a resource of degree at least 3, we perform the following operation. Root the tree containing such a resource at an arbitrary player. Consider a resource $j$ of degree at least 3 such that the subtree rooted at this resource contains only resources of degree 2. Because this resource must have at least 2 children in the tree $i_1,i_2,\ldots$ (which are players) and because 
\begin{equation*}
    \sum_{i\in P} \sum_{C : j\in C} x^*_{i,C}\leq 1,
\end{equation*}
it must be that one of the children (say $i_1$) satisfies $x^*_{i_1,\{j\}}\leq 1/2$. We then delete the edge $(j,i_1)$ in the tree and set $x^*_{i_1,\{j\}}$ to 0. 

\item Every resource now has degree exactly 2. We form a cluster for each tree in the forest. The cluster will contain the players and fat resources in the tree. We note that in every tree, only the player at the root lost at most $1/2$ unit of a fat resource by the previous step in the construction. By the degree property of resources and because the graph contains no cycle, it must be that in each cluster $K$ we have $|R(K)|=|P(K)|-1$ where $|R(K)|$ is the number of resources in the cluster and $|P(K)|$ the number of players. Because each resource is assigned at most once, and because only one player in the cluster lost at most $1/2$ unit of a fat resource, it must be that the cumulative amount of thin configurations assigned to players in $K$ is at least 
\begin{equation*}
    |P(K)|-|R(K)|-1/2=1/2.
\end{equation*}
This gives the second property of the lemma. For the third property, notice that for any choice of player $i\in K$, we can root the tree corresponding to the cluster $K$ at the player $i$ and assign all the fat resources in $K$ to their only child in the tree (they all have degree 2). This gives the third property of the lemma.

As each of these steps individually maintained maintained a congestion of at most $1$ on every thin resource, we indeed get a new solution $x'$ and the associated clusters with the required properties.
\end{enumerate}
\end{proof}

Lemma~\ref{lem:clusters} implies that
for each cluster we need to cover only one player
with a thin configuration. Then the remaining players
can be covered with fat resources.
We will now replace $x'$ by a solution $x''$
which takes slightly worse configurations
$\C_t(i, T^*/5)$,
but satisfies (2) in Lemma~\ref{lem:clusters}
with $2$ instead of $1/2$.
This can be achieved by splitting each
configuration $C\in \C_t(i, T^*)$
in $4$ disjoint parts $C_1, C_2, C_3, C_4\in \C_t(i, T^* / 5)$.
Let $C_1 \subseteq C$ with $f(C_1) \ge T^* / 5$
minimal in the sense that $f(C_1 \setminus \{j\}) < T^* / 5$
for all $j\in C_1$.
Let $j_1 \in C_1$. By submodularity and because $j_1$ is
thin it holds that
\begin{equation*}
    f(C \setminus C_1) \ge f(C) - f(C_1 \setminus \{j_1\}) - f(\{j_1\}) \ge 4T^* / 5 - T^*/100 .
\end{equation*}
Hence, in the same way we can select $C_2\subseteq C\setminus C_1$, $C_3\subseteq C\setminus (C_1 \cup C_2)$ and
$C_4\subseteq C\setminus (C_1 \cup C_2 \cup C_3)$.
We now augment $x'$ to $x''$ by initializing $x''$
with $0$ and then for each $i$ and $C\in\C(i,T^*)$
increasing $x''_{i, C_1}$, $x''_{i, C_2}$, $x''_{i, C_3}$, and $x''_{i, C_4}$ by $x'_{i, C}$.
Here $C_1, C_2, C_3, C_4 \in \C(i, T^* / 5)$ are the configurations derived from $C$ by splitting it as
described above.

Finally, we sample for each cluster some $\ell \geq 12\log(n)$ many
configurations with the distribution of $x''$ to
obtain the statement of Lemma~\ref{lem:config-sample} which we restate for convenience.
\begin{customthm}{\ref{lem:config-sample}} (restated)
  Let $\ell \ge 12\log (n)$.
  Given a solution of value $T^*$ for the configuration LP
  in randomized polynomial time we can find a partition of the players into clusters $K_1\cup\cdots \cup K_k\cup Q = P$ and multisets of configurations
  $\C_h \subseteq \bigcup_{i\in K_h} \C_T(i, T^*/5)$, $h=1,\dotsc,k$, such that
  \begin{enumerate}
      \item $|\C_h| = \ell$ for all $h=1,\dotsc,k$ and
      \item Each small resource appears in at most $\ell$ configurations of $\bigcup_h \C_h$.
    \item given any $i_1\in K_1, i_2\in K_2,\dotsc,i_k\in K_k$
    there is a matching of fat resources to players
    $P\setminus\{i_1,\dotsc,i_k\}$ such that each of these players $i$ gets a unique fat resource $j\in\Gamma_i$.
  \end{enumerate}
\end{customthm}
\begin{proof}
  We start with the clusters obtained with Lemma~\ref{lem:clusters} and the solution $x''$ described above. Recall that
  \begin{equation*}
      \sum_{i\in K_h} \sum_{C\in C_t(i,T^*/5)} x''_{i,C} \geq 2
  \end{equation*}
  for each cluster $K_h$. We assume w.l.o.g.\ that equality holds
  by reducing some variables $x''_{i, C}$. Clearly then each resource
  is still contained in at most one configuration in total.

  For each cluster $K_h$, we sample a configuration that contains a player in this cluster according to the probability distribution given by the values $\{x''_{i,C}/2 \}_{i\in K_h, C\in \C_t (i,T^*/5)}$. 
  By the assumption of equality stated above
  this indeed defines a probability distribution. We repeat this process $\ell$ times. We first note that for one iteration, each resource
  is in expectation contained in
  \begin{equation*}
      \sum_{i\in P} \sum_{C\in \C(i,T^*/5):j\in C} x''_{i,C}/2 \leq 1/2
  \end{equation*} selected configurations. Hence in expectation all the resource are contained in $\ell/2$ selected configurations after $\ell$ iterations. By a standard Chernoff bound (see Proposition \ref{chernoff}), we have that with probability at most 
  \begin{equation*}
      \exp \left( -\ell/6\right)\leq 1/n^2
  \end{equation*} a resource is contained in more than $\ell$ configurations. By a union bound, it holds that all resources are contained in at most $\ell$ selected configurations with high probability.
\end{proof}

\section{Omitted proofs from Section~\ref{sec:sequence}}\label{appendix_sequence}
\begin{customthm}{\ref{lma-size}}(restated)
Consider Random Experiment~\ref{exp:sequence}
with $\ell\geq 300.000\log^{3} (n)$.
For any $k\geq 0$ and any $C\in\C^{(\geq k)}$ we have 
  \begin{equation*}
      \frac{1}{2} \ell^{-k}|C| \le |\I_k \cap C| \le \frac{3}{2} \ell^{-k}|C|
  \end{equation*}
with probability at least $1-1/n^{10}$.
\end{customthm}
\begin{proof}
 The lemma trivially holds for $k=0$. 
 For $k>0$, by assumption $C\in\C^{(\geq k)}$ hence $|C|\geq \ell^{k+3}$. Since each resource of $\I=\I_0$ survives in $\I_k$ with probability $\ell^{-k}$ we clearly have that in expectation
 \begin{equation*}
     \mathbb E(|\I_k\cap C|) =  \ell^{-k}|C|
 \end{equation*}
 Hence the random variable $X=|\I_k\cap C|$ is a sum of independent variables of value either $0$ or $1$ and such that $\mathbb E (X)\geq \ell^3$. By a standard Chernoff bound (see Proposition \ref{cor:chernoff}), we get
 \begin{equation*}
     \mathbb P\left(X\notin \left[\frac{\mathbb{E}(X)}{2}, \frac{3\mathbb{E}(X)}{2}\right]\right) \leq 2 \exp \left(-\frac{\mathbb E(X)}{12} \right) \leq 2 \exp \left(-\frac{300.000\log^3 (n)}{12} \right) \leq  \frac{1}{n^{10}}
 \end{equation*}
 since by assumption $\ell \geq 300.000\log^3 (n)$.
\end{proof}

\begin{customthm}{\ref{lma-overlap-representative}}(restated)
Consider Random Experiment~\ref{exp:sequence}
with $\ell\geq 300.000\log^{3} (n)$.
For any $k\geq 0$ and any $C\in\C^{(\geq k)}$ we have 
  \begin{equation*}
      \sum_{C'\in \C^{(k)}} |C'\cap C\cap \I_k| \leq \frac{10}{\ell^{k}} \left(|C|+\sum_{C'\in \C^{(k)}} |C'\cap C| \right)
  \end{equation*}
with probability at least $1-1/n^{10}$.
\end{customthm}
\begin{proof} The expected value of the random variable $X=\sum_{C'\in \C^{(k)}} |C'\cap C\cap \I_k|$ is 
\begin{equation*}
    \mathbb E(X) = \frac{1}{\ell^k} \sum_{C'\in \C^{(k)}} |C'\cap C|.
\end{equation*}
Since each resource is in at most $\ell$ configurations, $X$ is a sum of independent random variables that take value in a range $[0,\ell]$. Then by a standard Chernoff bound (see Proposition \ref{cor:chernoff}), we get

\begin{equation*}
    \mathbb P\left(X\ge  10  \left(\frac{|C|}{\ell^k} + \mathbb E(X)\right) \right) \leq \exp\left(-\frac{3|C|}{\ell^{k+1}}\right) \leq \frac{1}{n^{10}} ,
\end{equation*}
since by assumption, $|C|\geq \ell^{k+3}$ and $\ell \geq 300.000\log ^3(n)$.

\end{proof}

We finish by the proof of the last property. As mentioned in the main body of the paper, this statement is a generalization of some ideas that already appeared in \cite{BansalSrividenko}. However, in \cite{BansalSrividenko}, the situation is simpler since they need to sample down the resource set only once (i.e. there are only two sets $R_1\subseteq R$ and not a full hierarchy of resource sets $R_d\subseteq R_{d-1}\subseteq \cdots \subseteq R_1 \subseteq R$). Given the resource set $R_1$, they want to select configurations and give to each selected configuration $K$ all of its resource set $|K\cap R_1|$ so that no resource is assigned too many times. In our case the situation is also more complex than that since at every step the selected configurations receive only a fraction of their current resource set. Nevertheless, we extend the ideas of Bansal and Srividenko to our more general setting. We recall the main statement before proceeding to its proof.
\begin{customthm}{\ref{lma-good-solution}}(restated)
Consider Random Experiment~\ref{exp:sequence}
with $\ell\geq 300.000\log^{3} (n)$. Fix $k\geq 0$. Conditioned on the event that the bounds in Lemma~\ref{lma-size} hold
for $k$, then with probability at least $1 - 1/n^{10}$
the following holds for all
$\F\subseteq \C^{(\geq k+1)}$, $\alpha:\F \rightarrow \mathbb N$, and $\gamma \in\mathbb N$ such that $\ell^3/1000\leq \alpha(C) \leq n $ for all $C\in\F$ and
$\gamma \in \{1,\dotsc,\ell\}$:
If there is a $(\alpha,\gamma)$-good assignment of $\I_{k+1}$ to $\F$, then there is a $(\alpha',\gamma)$-good assignment of $\I_k$ to $\F$ where
\begin{equation}\label{property:3}
    \alpha'(C) \ge \ell \left(1-\frac{1}{\log (n)} \right) \alpha(C)
\end{equation}
for all $C\in\F$.
Moreover, this assignment can be found in polynomial time.
\end{customthm}

We first provide the definitions of a flow network that allows us to state a clean condition whether a good assignment of resources exists or not. We then provide the high probability statements that imply the lemma.

\begin{figure}[]
    \centering
    \includegraphics[scale=1.15]{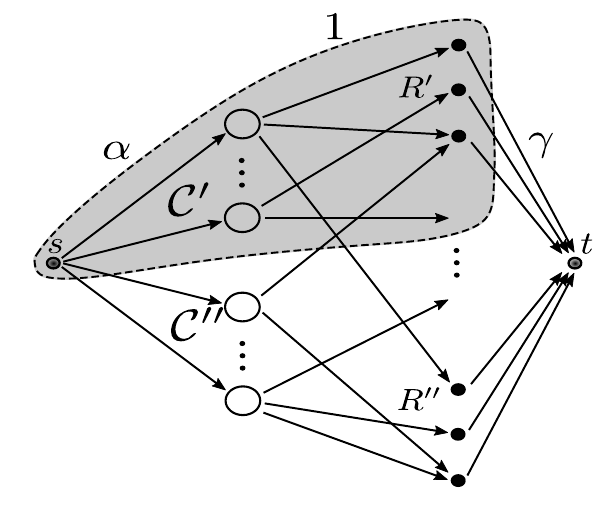}
    \caption{The directed network and an $s$-$t$ cut}
    \label{fig:network_flow}
\end{figure}

For any subset of configurations $\mathcal F \subseteq \C^{(\geq k+1)}$, resource set $\I_k$, $\alpha:\F \rightarrow \mathbb N$, and any integer $\gamma$, consider the following directed network (denoted by $\mathcal N (\mathcal F, \I_k, \alpha,\gamma)$). Create a vertex for each configuration in $\mathcal F$ as well as a vertex for each resource. Add a source $s$ and sink $t$. Then add a directed arc from $s$ to the vertex 
$C\in\mathcal F$ with capacity $\alpha(C)$. For every pair of a configuration $C$ and a resource $i$ such that $i\in C$ add a directed arc from $C$ to $i$ with capacity $1$. Finally, add a directed arc from every resource to the sink of capacity $\gamma$. See Figure \ref{fig:network_flow} for an illustration. 

We denote by 
\begin{equation*}
    \textrm{maxflow}\left(\mathcal N (\mathcal F, \I_k, \alpha,\gamma)\right)
\end{equation*}
the value of the maximum $s$-$t$ flow in $\mathcal N (\mathcal F, \I_k, \alpha,\gamma)$.

Before delving into the technical lemmas, we provide a brief road map for the proof. First, we argue that for any subset of configurations, in the two networks induced on this subset and the consecutive resource sets (which are $\I_k$ and $\I_{k+1}$), the value of the maximum flow differs by approximately a factor $\ell$ (this is Lemma \ref{lem:flow_conservation} stated below). Then by a union bound over all possible subsets of configurations, we say that the above argument consecutively holds with good probability. This helps us conclude that a good assignment of the resource set $\I_{k+1}$ implies that there is a good assignment of the resource set $\I_k$. Notice that if one does not have the above argument with respect to all subsets of configurations at once, it is not necessary that a good assignment of resources must exist. In particular, we need Lemma \ref{lem:flow_black_box} to show that if on \textit{all} subsets of configurations the maximum flow is multiplied by \textit{approximately} $\ell$ when we expand the resource set from $\I_{k+1}$ to $\I_k$, then an $(\alpha,\gamma)$-good assignment of $\I_{k+1}$ implies an $(\alpha',\gamma)$-good assignment of $\I_k$, where $\alpha'$ is almost equal to $\ell \alpha$.

\begin{lemma}\label{lem:flow_black_box}
Let $\mathcal F$ be a set of configurations,
$\I' \subseteq \I$, $\alpha:\F \rightarrow \mathbb N$ a set of resources, $\gamma\in\mathbb N$, and $\epsilon \ge 0$.
Define
\begin{equation*}
    \alpha'(C) = \lfloor (1 - \epsilon) \alpha(C) \rfloor .
\end{equation*}
There is an $(\alpha', \gamma)$-good assignment of $\I'$ to $\F$ if and only if for every $\F' \subseteq \F$, the maximum flow in the network $\mathcal N (\mathcal F', \I', \alpha,\gamma)$ is of value at least $\sum_{C\in \F'}\alpha'(C)$. Moreover, this assignment can be found in polynomial time.
\end{lemma}

\begin{proof}
First assume there is such an $(\alpha', \gamma)$-good assignment. Then send a flow of $\alpha'(C)$ from $s$ to each $C\in \F$. If resource $i$ is assigned to $C$, send a flow of $1$ from $C$ to $i$. Finally ensure that flow is preserved at every vertex corresponding to a resource by sending the correct amount of flow to $t$. Since no resource is taken more than $\gamma$ times, this flow is feasible. 

We prove the other direction by contradiction. Denote by $\N$ the network $\N(\F,\I',\alpha',\gamma)$.
If there is no good assignment satisfying the condition of the lemma then the maximum flow in $\N$ must be strictly less than $\sum_{C\in \F}\alpha'(C)$ (otherwise consider the maximum flow, which can be taken to be integral, and give to every configuration $C$ all the resources to which they send a flow of $1$). Then by the max-flow min-cut theorem, there exists an $s$-$t$ cut $S$ that has value strictly less than $\sum_{C\in \F}\alpha'(C)$. Let $\mathcal C'$ be the set of configurations on the side of the source in $S$. Notice that $\mathcal C'$ cannot be empty by assumption on the value of the cut.

Consider the induced network $\mathcal N (\mathcal C', \I', \alpha',\gamma)$ and the cut $S$ in it. It has a value strictly lower than $\sum_{C\in \mathcal C'} \alpha'(C)$. This, in turn implies that the cut $S$ in $\mathcal N (\mathcal C', \I', \alpha,\gamma)$ has a value strictly lower than $\sum_{C\in \mathcal C'} \alpha'(C)$, since this cut does not contain any edge from the source $s$ to some configuration. Hence the maximum flow in $\mathcal N (\mathcal C', \I', \alpha,\gamma)$ has a value strictly less than $\sum_{C\in \mathcal C'} \alpha'(C)$,
a contradiction to the assumption in the premise. 
\end{proof}

\begin{lemma}
\label{lem:flow_conservation}
Let $\mathcal F\subseteq \mathcal C^{\geq (k+1)}$, $\alpha:\F \rightarrow \mathbb N$ such that $\ell^3/1000 \leq \alpha(C) \leq n$ for all $C\in\F$, and $1 \le \gamma \le \ell$. Denote by $\mathcal N$ the network $\mathcal N (\mathcal F, \I_k, \ell \cdot \alpha,\gamma)$ and by $\Tilde{\mathcal{N}}$ the network $\mathcal N (\mathcal F, \I_{k+1},\alpha,\gamma)$. Then
 \begin{equation*}
     \mathrm{maxflow}\left(\mathcal{N}\right)\geq \frac{\ell}{1+0.5/\log (n)}  \mathrm{maxflow}\left(\Tilde{\mathcal{N}}\right)
 \end{equation*}
 with probability at least $1-1/(n\ell)^{20|\mathcal F|}$.
\end{lemma}
\begin{proof}
We use the max-flow min-cut theorem that asserts that the value of the maximum flow in a network is equal to the value of the minimum $s$-$t$ cut in the network. Consider a minimum cut $S$ of network $\mathcal N$ with $s\in S$ and $t\notin S$. Denote by $c(S)$ the value of the cut. We will argue that with high probability this cut induces a cut of value at most $c(S) / \ell \cdot (1+0.5/\log(n))$
in the network $\Tilde{\mathcal N}$.
This directly implies the lemma.

Denote by $\mathcal C'$ the set of configurations of $\F$ that are in $S$, i.e., on the source side of the cut, and $\mathcal C''=\mathcal{F}\setminus \mathcal C'$.
Similarly consider $\I'$ the set of resources in the $s$ side of the cut and $\I''= \I_k\setminus \I'$. With a similar notation, we denote $\Tilde \I' = \I'\cap \I_{k+1}$ the set of resources of $\I'$ surviving in $\I_{k+1}$; and $\Tilde \I'' = \I''\cap \I_{k+1}$. Finally, denote by $\Tilde S$ the cut in $\Tilde{\mathcal N}$ obtained by removing resources of $R'$ that do not survive in $\I_{k+1}$ from $S$, i.e.,
$\Tilde S = \{s\}\cup \mathcal C' \cup \I'$.
The value of the cut $S$ of $\mathcal N$ is
\begin{equation*}
    c(S) = \sum_{C\in \mathcal C''} \ell \cdot \alpha(C) + e(\mathcal C',\I'')+ \gamma  |\I'|
\end{equation*}
where $e(X,Y)$ denotes the number of edges from $X$ to $Y$.
The value of the cut $\Tilde S$ in $\Tilde{\mathcal N}$ is
\begin{equation*}
    c( \Tilde S) = \sum_{C\in \mathcal C''} \alpha(C) + e(\mathcal C',\Tilde \I'')+ \gamma  |\Tilde \I'|
\end{equation*}
We claim the following properties.
\begin{claim}
\label{cla:size_configurations}
For every $C\in \mathcal F$, the outdegree of the vertex corresponding to $C$ in $\mathcal N$ is at least $\ell^4/2$.
\end{claim}
Since $C\in \C^{(\geq k+1)}$ and by Lemma \ref{lma-size}, we clearly have that $|C\cap \I_k|\geq \ell^4/2$.
\begin{claim}
\label{cla:size_cut}
It holds that
\begin{equation*}
 c(S)\geq \frac{|\F| \ell^3}{1000} .
\end{equation*}
\end{claim}
We have by assumption on $\alpha(C)$
\begin{multline*}
    c(S) = \sum_{C\in \mathcal C''} \ell \cdot \alpha(C) + e(\mathcal C',\I'')+ \gamma |\I'|
    \geq \sum_{C\in \mathcal C''} \frac{\ell^3}{1000} + e(\mathcal C',\I'')+ \gamma |\I'|\\
    \geq \frac{|\mathcal C''|\ell^3}{1000} + e(\mathcal C',\I'')+ \gamma |\I'|
\end{multline*}
Now consider the case where $e(\mathcal C',\I'')\leq |\mathcal C'|\ell^3 / 1000$.
Since each vertex in $\mathcal C'$ has outdegree at least $\ell^4/2$ in the network $\mathcal N$ (by Claim~\ref{cla:size_configurations}) it must be that $e(\mathcal C',\I')\geq |\mathcal C'|\ell^4 / 2 - |\mathcal C'|\ell^3 / 1000 > |\mathcal C'|\ell^4 / 3$.
Using that each vertex in $\I'$ has indegree at most $\ell$ (each resource is in at most $\ell$ configurations), this implies
$|\I'|\geq |\mathcal C'|\ell^3 / 3$. Since $\gamma \geq 1$ we have in all cases that $e(\mathcal C',\I'')+ \gamma  |\I'|\geq |\mathcal C'|\ell^3 / 1000$. Hence 
\begin{equation*}
    c(S) \geq \frac{|\mathcal C''|\ell^3}{1000} + \frac{|\mathcal C'|\ell^3}{1000} =  \frac{|\F| \ell^3}{1000} .
\end{equation*}
This proves Claim~\ref{cla:size_cut}.
We can now finish the proof of the lemma.
Denote by $X$ the value of the random variable $e(\mathcal C',\Tilde{\I''})+ \gamma  |\Tilde{\I'}|$. We have that 
\begin{equation*}
    \mathbb E[X] = \frac{1}{\ell}(e(\mathcal C',\I'')+ \gamma  |\I'|).
\end{equation*}
Moreover, $X$ can be written as a sum of independent variables in the range $[0, \ell]$ since each vertex is in at most $\ell$ configurations and $\gamma \le \ell$ by assumption. By a Chernoff bound (see Proposition \ref{cor:chernoff}) with
\begin{equation*}
    \delta = \frac{0.5 c(S)}{\log(n) \cdot (c(S)-\sum_{C\in \mathcal C''} \alpha(C))} \geq \frac{0.5}{\log(n)}
\end{equation*} 
we have that
\begin{multline*}
    \mathbb P\left(X\geq \mathbb E(X)+\frac{0.5 c(S)}{\ell\log(n)}\right) 
    \leq \exp\left(-\frac{\min\{\delta,\delta^2\}\mathbb E(X)}{3\ell} \right) \\
    \leq \exp\left(-\frac{c(S)}{12\ell^2\log^2 (n)} \right)
    \leq \exp\left(-\frac{|\mathcal F|\ell^3}{12.000\ell^2\log^2 (n)} \right)
    \leq \frac{1}{(n\ell)^{20|\F|}} ,
\end{multline*}
where the third inequality comes from Claim~\ref{cla:size_cut} and the last one from the assumption that $\ell\geq 300.000\log^{3}(n)$.
Hence with probability at least $1-1/(n\ell)^{20|\F|}$, we have that 
\begin{equation*}
    c( \Tilde S) = \sum_{C\in \mathcal C''} \alpha(C) + e(\mathcal C',\Tilde \I'')+ \gamma |\Tilde \I'| \leq \frac{1}{\ell}c(S)+\frac{0.5}{\ell \log (n)}c(S) .\qedhere
\end{equation*}
\end{proof}
We are now ready to prove Lemma~\ref{lma-good-solution}.
Note that Lemma \ref{lem:flow_conservation} holds with probability at least $1-1/(n\ell)^{20|\F|}$.
Given the resource set $\I_k$ and a cardinality $s = |\F|$ there are $O((n\ell)^{2s})$ ways of defining a network satisfying the conditions from Lemma~\ref{lem:flow_conservation} ($(m\ell)^s\le (n\ell)^s$ choices of $\F$, $n^{s}$ choices for $\alpha$ and $\ell$ choices for $\gamma$). By a union bound, we can assume that the properties of Lemma~\ref{lem:flow_conservation} hold for every possible network with probability at least $1 - 1/n^{10}$.
Assume now there is a $(\alpha,\gamma)$-good assignment of $\I_{k+1}$ to some family $\F$. Then by Lemma~\ref{lem:flow_black_box} the $\mathrm{maxflow}(\N(\F',\I_{k+1}, \alpha,\gamma))$ is exactly $\sum_{C\in \F'}\alpha(C)$ for any $\F'\subseteq \F$. By Lemma~\ref{lem:flow_conservation}, this implies that $\mathrm{maxflow}(\N(\F',\I_{k}, \ell \cdot \alpha,\gamma))$ is at least $\ell/(1+0.5/\log(n)) \sum_{C\in \F'}\alpha(C)$. By Lemma \ref{lem:flow_black_box}, this implies a $(\alpha',\gamma)$-good assignment from $\I_k$ to $\F$, where
\begin{equation*}
    \alpha'(C) = \lfloor\ell/(1+0.5/\log(n))\rfloor \alpha(C) \ge \ell / (1 + 1/\log(n)) \alpha(C) \geq \ell(1 - 1/\log(n)) \alpha(C).
\end{equation*}

\section{Omitted proofs from Section~\ref{sec:reconstruction}}\label{appendix_reconstruct}
\begin{customcla}{\ref{cla:reconstruct}}(restated)
For any $k\geq 0$, any $0\leq j\leq k$, and any $C\in \K^{(k)}$
\begin{equation*}
    \sum_{j\leq h\leq k}\sum_{K\in \K^{(h)}} \ell^{h}|K\cap C \cap \I_h| \leq 2000\frac{d+\ell}{\ell}\log (\ell) |C|.
\end{equation*}
\end{customcla}

\begin{proof}[Proof of Claim \ref{cla:reconstruct}]
By Lemma~\ref{lma:main-LLL} we have that 
\begin{equation*}
    \sum_{j\leq h\leq k}\sum_{K\in \K^{(h)}} \ell^{h}|K\cap C \cap \I_h| \leq \frac{1}{\ell} \sum_{j\leq h\leq k}\sum_{C'\in \C^{(h)}} \ell^{h}|C'\cap C \cap \I_h| + 1000\frac{d+\ell}{\ell}\log (\ell) |C|.
\end{equation*}
Furthermore, by Lemma \ref{lma-overlap-representative}, we get 
\begin{equation*}
    \sum_{C'\in \C^{(h)}} \ell^{h}|C'\cap C \cap \I_h| \leq \ell^{h}\frac{10}{\ell^h}\left(|C|+\sum_{C'\in \C^{(h)}} |C'\cap C| \right).
\end{equation*}
Finally note that each resource appears in at most $\ell$ configurations, hence
\begin{equation*}
    \sum_{j\leq h\leq k}\sum_{C'\in \C^{(h)}} |C'\cap C| \leq \ell |C|.
\end{equation*}
Putting everything together we conclude
\begin{align*}
    \sum_{j\leq h\leq k}\sum_{K\in \K^{(h)}} \ell^{h}|K\cap C \cap \I_h| &\leq \frac{1}{\ell} \sum_{j\leq h\leq k}\sum_{C'\in \C^{(h)}} \ell^{h}|C'\cap C \cap \I_h| + 1000\frac{d+\ell}{\ell}\log (\ell) |C| \\
    &\leq \frac{1}{\ell} \sum_{j\leq h\leq k}10\left( |C|+\sum_{C'\in \C^{(h)}}|C'\cap C|\right) + 1000\frac{d+\ell}{\ell}\log (\ell) |C|\\
    &\leq \frac{k-j}{\ell}10|C|+10|C|+1000\frac{d+\ell}{\ell}\log (\ell) |C|\\
    &\leq 20|C|+1000\frac{d+\ell}{\ell}\log (\ell) |C|\\
    &\leq 2000\frac{d+\ell}{\ell}\log (\ell) |C|.\qedhere
\end{align*}
\end{proof}

\begin{customcla}{\ref{cla:reconstruct_mu}}(restated) For any $C\in \K^{(\geq j)}$,
\begin{equation*}
    \frac{1}{\gamma^2}\sum_{K\in \K^{(j-1)}}|K\cap C \cap \I_{j-1}\cap O|\leq \mu \leq \frac{2}{\gamma} \sum_{K\in \K^{(j-1)}}|K\cap C \cap \I_{j-1}\cap O|.
\end{equation*}
\end{customcla}
\begin{proof}[Proof of Claim \ref{cla:reconstruct_mu}]
Note that we can write 
\begin{equation*}
    \mu = \sum_{i\in O\cap C} \frac{a_i+b_i-\gamma}{b_i} \leq \max_{i\in O\cap C}\left\lbrace \frac{a_i+b_i-\gamma}{a_ib_i} \right\rbrace \sum_{K\in \K^{(j-1)}}|K\cap C \cap \I_{j-1}\cap O|.
\end{equation*}
The reason for this is that each resource $i$ accounts for an expected loss of $(a_i+b_i-\gamma)/b_i$ while it is counted $a_i$ times in the sum 
\begin{equation*}
    \sum_{K\in \K^{(j-1)}}|K\cap C \cap \I_{j-1}\cap O|.
\end{equation*}
Similarly,
\begin{equation*}
    \mu = \sum_{i\in O\cap C} \frac{a_i+b_i-\gamma}{b_i} \geq \min_{i\in O\cap C}\left\lbrace \frac{a_i+b_i-\gamma}{a_ib_i} \right\rbrace \sum_{K\in \K^{(j-1)}}|K\cap C \cap \I_{j-1}\cap O|.
\end{equation*}
Note that by assumption we have that $a_i+b_i>\gamma$. This implies that either $a_i$ or $b_i$ is greater than $\gamma/2$. Assume w.l.o.g. that $a_i\geq \gamma/2$. Since by assumption $a_i\leq \gamma$ we have that 
\begin{equation*}
    \frac{a_i+b_i-\gamma}{a_ib_i}\leq \frac{b_i}{a_ib_i} =\frac{1}{a_i} \leq \frac{2}{\gamma}.
\end{equation*}
In the same manner, since $a_i+b_i>\gamma$ and that $a_i,b_i\leq \gamma$, we can write
\begin{equation*}
    \frac{a_i+b_i-\gamma}{a_ib_i}\geq \frac{1}{a_ib_i} \geq \frac{1}{\gamma^2}.
\end{equation*}
We therefore get the following bounds
\begin{equation*}
    \frac{1}{\gamma^2}\sum_{K\in \K^{(j-1)}}|K\cap C \cap \I_{j-1}\cap O|\leq \mu \leq \frac{2}{\gamma} \sum_{K\in \K^{(j-1)}}|K\cap C \cap \I_{j-1}\cap O|,
\end{equation*}
which is what we wanted to prove.
\end{proof}

\section{Omitted proofs from Section~\ref{sec:reduction santa claus}}\label{appendix:reduction}
\subsection{From matchings to Santa Claus}
The idea in this reduction is to replace each player by a set of players, one for each of the $t$ configuration containing him.
These players will share together $t-1$ large new resources, but to satisfy all, one of them has to get other resources, which
are the original resources in the corresponding configuration.
\begin{description}
\item[Players.]
For every vertex $v \in P$, and every hyperedge $C \in \C$ that $v$ belongs to, we create a player $p_{v,C}$ in the Santa Claus instance. 
\item[Resources.]
For every vertex $u \in \I$, create a resource $r_{u}$ in the Santa Claus instance. 
For any vertex $v \in P$ such that it belongs to $t$ edges in $\C$, create $t-1$ resources $r_{v,1}, r_{v,2}, \ldots, r_{v,t-1} $. 
\item[Values.]
For any resource $r_{u}$ for some $u \in \I$ and any player $p_{v,C}$ for some $C \in \C$, the resource has a value $\frac{1}{|C|-1}$ if $u \in C$, otherwise it has value $0$. Any resource $r_{v,i}$ for some $v \in P$ and $i \in \mathbb N$, has value $1$ for any player $p_{v,C}$ for some $C \in \C$ and $0$ to all other players. 
\end{description}
It is easy to see that given an $\alpha$-relaxed matching in the original instance, one can construct an $\alpha$-approximate solution for the Santa Claus instance.

For the other direction, notice that for each $v \in P$, there exists a player $p_{v,C}$ for some $C \in \C$, such that it gets resources only of the type $r_{u}$. One can simply assign the resource $u \in \I$ to the player $v$ for any resource $r_{u}$ assigned to $p_{v,C}$.

\subsection{From Santa Claus to matchings}
This subsection is devoted to the proof of Theorem \ref{thm:reduction}.

\begin{proof}
We write $(\log)^k(n) = \underbrace{\log \cdots \log}_{\times k}(n)$ and $(\log)^0(n) = n$.   

\paragraph*{Construction.}
We describe how to construct a hypergraph matching instance from a Santa Claus instance in four steps by reducing to the following more and more special cases.

\paragraph{(1) Geometric grouping.} In this step, given arbitrary $v_{ij}$, we reduce it to an instance such that $\OPT = 1$ and for each $i, j$ we have $v_{ij} = 2^{-k}$ for some integer $k$ and $1/(2n) < v_{ij} \le 1$.
    This step follows easily from guessing $\OPT$, rounding down the sizes, and omitting all small elements in a solution.
\paragraph{(2) Reduction to O(log*(n)) size ranges.} Next, we reduce to an instance such that for each player $i$ there is some $k \le \log^*(2n)$ such that for each resource $j$, $v_{ij}\in\{0, 1\}$ or $1/(\log)^k(2n) < v_{ij} \le 1/(\log)^{k+1}(2n)$. We explain this step below.
    
    Each player and resource is copied to the new instance.
However, we will also add auxiliary players and resources.
Let $i$ be a player.
In the optimal solution there is some
$0 \le k \le \log^*(2n)$ such that the values of all resources
$j$ with $1/(\log)^k(2n) < v_{ij} \le 1/(\log)^{k+1}(2n)$ assigned
to player $i$ sum up to at least $1/\log^*(2n)$.
Hence, we create $\log^*(2n)$ auxiliary players which correspond to
each $k$ and each of which share an resource with the original player that
has value $1$ for both.
The original player needs to get one of these resources, which means
one of the auxiliary players needs to get a significant value from
the resources with $1/(\log)^k(2n) < v_{ij} \le 1/(\log)^{k+1}(2n)$.
This reduction loses a factor of at most $\log^*(2n)$.
Hence, $\OPT \geq 1/\log^*(2n)$. 

\paragraph{(3) Reduction to 3 sizes.} We further reduce to an instance such that for each player $i$ there is some value $v_i$ such that for each resource $j$, $v_{ij}\in\{0, v_i, 1\}$. 

Let $i$ be some player who has only resources of value $v_{ij}\in\{0,1\}$ or
$1/(\log)^k(2n) < v_{ij} \le 1/(\log)^{k+1}(2n)$ for some integer $k$. 
There are at most $\log((\log)^k(2n)) \leq (\log)^{k+1}(2n)$ distinct values of
the latter kind. The idea is to assign bundles of resources of value $0.5/\left( \log^*(2n)(\log)^{k+1}(2n) \right)$ to the player $i$. 

Fix a resource value $s$ such that $1/(\log)^k(2n) <s\le 1/(\log)^{k+1}(2n)$. We denote by $\I_s$ the set of resources $j$ such that $v_{ij}=s$.

We define the integer
\begin{equation*}
    b=\left\lceil \frac{0.5}{s\log^*(2n)(\log)^{k+1}(2n)}\right\rceil
\end{equation*} which is the number of resources of value $s$ that are needed to make a bundle of total value at least $0.5/\left( \log^*(2n)(\log)^{k+1}(2n) \right)$. We remark that if $s>0.5/\left( \log^*(2n)(\log)^{k+1}(2n) \right)$ we have $b=1$. However, since $s\leq 1/(\log)^{k+1}(2n)$, the value of a bundle never exceeds $1/(\log)^{k+1}(2n)$ in the instance of step (2).

Then we create
\begin{equation*}
    \left\lfloor |\I_s|/b\right\rfloor
\end{equation*} auxiliary players $i_1,i_2,\ldots $ and auxiliary resources $j_1,j_2,\ldots$ (note that we create 0 player and resource if $|\I_s|<b$).

Each auxiliary player $i_\ell$ shares resource $j_\ell$ with player $i$. This resource has value $2/\left( \log^*(2n)(\log)^{k+1}(2n) \right)$ for player $i$ and value $1$ for player $i_\ell$. Then for all resources $j\in R_s$, we set $v_{ij}=0$ and 
\begin{equation*}
    v_{i_\ell j}=\frac{1}{(\log^*(2n))^2b}
\end{equation*} for any auxiliary player $i_\ell$ that was created.

We see that we are now in the case where for each player $i$, there exists some $v_i$ such that $v_{ij}\in \{0,v_i,1\}$ for all resources $j$. We claim the following.
\begin{claim}
\label{cla:reduction_3_OPT}
In the instance created at step (3), we have that $\OPT\geq 1/(\log^*(2n))^2$.
\end{claim}
\begin{proof}
To see this, take an assignment of resources to player that gives $1/\log^*(2n)$ value to every player in the instance obtained at the end of step (2). Define $\I_i$ to be the set of resources assigned to player $i$ in this solution. Either $\I_i$ contains a resource of value $1$ or only resources that are in a range $(1/(\log)^k(2n),1/(\log)^{k+1}(2n)]$ for some integer $k$. In the first case, nothing needs to be done as the resource $j$ of value $1$ assigned to $i$ still satisfies $v_{ij}=1$ in the new instance. Hence we assign $j$ to $i$ and all auxiliary players created for player $i$ get their auxiliary resource of value 1. 

In the second case, fix a resource value $s$. Let $\I_{i,s}$ be the set of resources assigned to $i$ for which $v_{ij}=s$ and $b$ defined as before. We select $\left\lfloor |\I_{i,s}|/b \right\rfloor$ auxiliary players to receive $b$ resources from $\I_{i,s}$ and player $i$ takes the corresponding auxiliary resources. The remaining auxiliary players of the corresponding value take their auxiliary resource. 

Doing this, we ensure that all auxiliary players receive either a value of 1 (by taking the auxiliary resource) or $1/(\log^*(2n))^2$ by taking resources assigned to $i$ in the instance of step (2). Moreover, we claim that $i$ receives a total value of at least $1/(\log^*(2n))^2$. To see this, we have $3$ cases depending on the value of $b$ and $\left\lfloor |\I_{i,s}|/b \right\rfloor$.

\begin{itemize}
    \item If $b=1$, then $\left\lfloor |\I_{i,s}|/b \right\rfloor=|\I_{i,s}|$. We note that the value of a bundle of $b$ resources of size $s$ never exceeds $1/(\log)^{k+1}(2n)$ in instance (2). Since each auxiliary resource represents a value of $2/\left( \log^*(2n)(\log)^{k+1}(2n) \right)$ to player $i$ in instance (3), it must be that player $i$ receives in instance (3) at least a $2/\log^*(2n)$ fraction of the value he would receive in instance (2).
    \item If $b>1$ and $\left\lfloor |\I_{i,s}|/b \right\rfloor>0$. Then we have that $\left\lfloor |\I_{i,s}|/b \right\rfloor\geq  |\I_{i,s}|/(2b)$. Since in this case we have $s<0.5/\left( \log^*(2n)(\log)^{k+1}(2n) \right)$ it must be that each bundle of $b$ resources of size $s$ represents a total value of at most $1/\left( \log^*(2n)(\log)^{k+1}(2n) \right)$. Since the value of auxiliary resources is twice this value and because $\left\lfloor |\I_{i,s}|/b \right\rfloor\geq  |\I_{i,s}|/(2b)$ it must be that in this case player $i$ receives in instance (3) at least the same value he would receive in instance (2).
    \item If $\left\lfloor |\I_{i,s}|/b \right\rfloor=0$, then player $i$ receives $0$ value from resources of this value. However, when we combine all the values $s$ for which $\left\lfloor |\I_{i,s}|/b \right\rfloor=0$, it represents to player $i$ in instance (2) a total value of at most 
    \begin{equation*}
        0.5/\left( \log^*(2n)(\log)^{k+1}(2n) \right)\cdot (\log)^{k+1}(2n) = 0.5/\log^*(2n)
    \end{equation*}
    since there are at most $(\log)^{k+1}(2n)$ different resource values.
\end{itemize}

Putting everything together, we see that in the first two cases, player $i$ receives at least a $2/\log^*(2n)$ fraction of the value he would receive in instance (2) and that he looses at total value of at most $0.5/\log^*(2n)$ in the third case. Since in instance (2) we have that $\OPT\geq 1/\log^*(2n)$ we see that in instance (3) player $i$ receives a value at least
\begin{equation*}
    (2/\log^*(2n))\cdot (1/\log^*(2n)-0.5/\log^*(2n))\geq 1/(\log^*(2n))^2.
\end{equation*}
\end{proof}

Finally, we also claim that it is easy to reconstruct an approximate solution to the instance obtained at step (1) from an approximate solution to the instance at step (3).
\begin{claim}\label{cla:reduction_3_to_1}
A $c$-approximate solution to the instance obtained at step (3) induces a $O((c\log^*(2n))^2)$-approximate solution to the instance obtained at step (1).
\end{claim}
\begin{proof}
To see this, note that a $c$-approximate solution must give at least $1/(c(\log^*(2n))^2)$ value to every player since $\OPT\geq 1/(\log^*(2n))^2$ (by Claim \ref{cla:reduction_3_OPT}). This means that each player $i$ either takes a resource of value 1 which has also value 1 for him in the instance at step (1) or he must take at total value of $1/(c(\log^*(2n))^2)$ in auxiliary resources and the corresponding auxiliary players must take bundles of resources that represent a value of at least 
\begin{equation*}
    0.5/\left( c\log^*(2n)(\log)^{k+1}(2n)\right)
\end{equation*}
for player $i$ in the instance at step (1). We simply assign all the resources appearing in these bundles to the player $i$ in the instance of step (1). Since the value of an auxiliary resource for player $i$ is $2/\left( \log^*(2n)(\log)^{k+1}(2n) \right)$ it must be that player $i$ takes at least 
\begin{equation*}
    \frac{1/(c(\log^*(2n))^2)}{2/\left( \log^*(2n)(\log)^{k+1}(2n) \right)} = \frac{(\log)^{k+1}(2n)}{2c\log^*(2n)}
\end{equation*} auxiliary resources. Since each auxiliary resource brings a value of 
\begin{equation*}
    0.5/\left( c\log^*(2n)(\log)^{k+1}(2n)\right)
\end{equation*} to player $i$ (in the instance at step (1)) then player $i$ receives in total a value of at least
\begin{equation*}
    \frac{1}{(2c\log^*(2n))^2}
\end{equation*} in the instance of step (1).
\end{proof}

Before the last step, we rescale the instance appropriately to get $\OPT=1$ (we keep the property that each player $i$ has 3 distinct sizes 0,1 and $v_i$). 

\paragraph{(4) Reduction to hypergraph matching.}
For each player create a vertex in $P$ and for
each resource create a vertex in $\I$.
For each player add one hyperedge for each resource he values at $1$ (containing $i$ and this resource).
Moreover, for every player $i$, add $1/v_i$ \textit{new} vertices to $P$ and
the same number of \textit{new} resources to $\I$. Pair these $1/v_i$ new vertices in $P$ and $\I$ together (one from $\I$ and one from $P$) and for each pair add a hyperedge containing these two vertices in the pair. Add another hyperedge for $i$ containing $i$ and all corresponding $1/v_i$ new vertices in $\I$. Finally, for each new vertex in $P$ and
each resource that $i$ values at $v_i$, add a hyperedge containing them. See Figure \ref{fig:reduction} for an illustration: New resources and players are marked as squares and hyperedges containing only 2 vertices are marked as simple edges.

\begin{figure}
    \centering
    \includegraphics[scale=1]{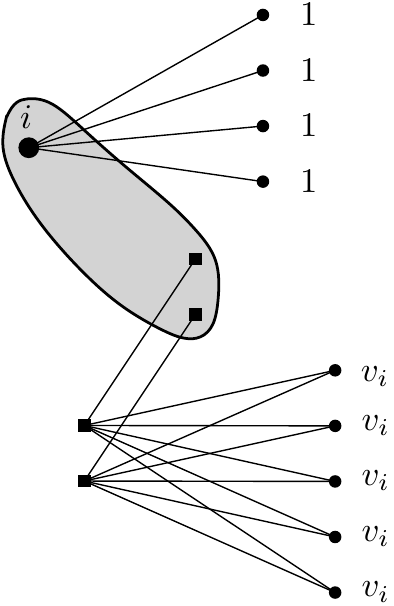}
    \caption{An example of the reduction to hypergraph matching for player $i$ with $v_i=1/2$.}
    \label{fig:reduction}
\end{figure}

We claim that there exists a $1$-relaxed perfect matching in this instance. Since $\OPT=1$ there is an assignment of resources to players such that every player gets a value $1$. If player $i$ takes one resource of value $1$, give to player $i$ the corresponding hyperedge and the resource in it in the hypergraph. All the new players get the new resource they are paired to. If player $i$ takes $1/v_i$ resources of value $v_i$, give to player $i$ in the hypergraph all the $1/v_i$ new resources contained in the new hyperedge. Then we give to each new player the hyperedge (and the resource in it) corresponding to a resource that is assigned to $i$ in instance from step (3). This is indeed a $1$-relaxed perfect matching.

\paragraph*{Correctness.} In the reduction we arrive at step (3) for which we prove that a $c$-approximate solution can be used to easily reconstruct a $O((c\log^*(2n))^2)$-approximate solution to the original instance (in Claim \ref{cla:reduction_3_to_1}). It remains to show that a $c$-relaxed perfect matching in the instance (4) induces a $c$-approximate solution to step (3). To see this, note that a $c$-relaxed perfect matching in the instance (4) either gives to player $i$ the resource in one hyperedge corresponding to a resource of value $1$ to player $i$ in instance (3). In that case we assign this resource to player $i$ in instance (3). Or it gives at least $1/(cv_i)$ new resources to player $i$. In this case, it must be that each new player paired to one of these resources takes one resource of value $v_i$ in instance (3). We give these resources to $i$ in instance (3). In this case $i$ receives a total value of $v_i/(cv_i)=1/c$ which ends the proof.

We finish by remarking that the size of our construction is indeed polynomial in the size of the original instance. This is clear for step (1). In step (2), only $O(\log^* (n))$ new players and items are created for each player in the original instance. In step (3), for each player $i$ and each resource size $v_{ij}$, at most a polynomial number of resources and players are created. As for the last step, $O(1/v_i)$ new resources and players are created for each player $i$ which is also polynomial since $v_i=\Omega (1/n)$. The number of hyperedges in the hypergraph is also clearly polynomial in the number of vertices in our construction.
\end{proof}

\end{document}